\documentclass[fleqn,usenatbib]{mnras}
\usepackage{newtxtext,newtxmath}
\usepackage{comment}

\usepackage[T1]{fontenc}
\usepackage{subfig}
\usepackage{float}
\DeclareRobustCommand{\VAN}[3]{#2}
\let\VANthebibliography\thebibliography
\def\thebibliography{\DeclareRobustCommand{\VAN}[3]{##3}\VANthebibliography}
\usepackage{graphicx}
\usepackage[tableposition=top]{caption}
\DeclareUnicodeCharacter{2212}{-}
\title{\centering{Beyond the Rings: Polar Ring Galaxy NGC 4262 and its Globular Cluster System}}
\author[Akhil Krishna R]{
Akhil Krishna R\thanks{E-mail: akhil.r@res.christuniversity.in}
, Sreeja S Kartha, and Blesson Mathew
\\
Department of Physics and Electronics, CHRIST (Deemed to be University), Bangalore 560029, India\\
}
\date{Accepted 2024 April 14. Received 2024 April 12; in original form 2024 March 23}

\pubyear{2024}

\begin{document}
\label{firstpage}
\pagerange{\pageref{firstpage}--\pageref{lastpage}}
\maketitle

\begin{abstract}
In the context of the hierarchical model of galaxy evolution, polar ring galaxies (PRGs) are considered the intermediate phase between ongoing mergers and quiescent galaxies. This study explores the globular cluster system (GCS) and its properties in the nearest PRG, NGC4262, serving as a pilot investigation to study GCS in nearby PRGs. We utilize wide and deep field observations of the CFHT as part of the NGVS to investigate the GCS of NGC4262. We presented the first optical image of NGC4262 with an optically faint ring component. The photometric analysis of the GCS displays a distinct color bimodality. We estimate the total number of GCs for NGC4262 to be 266$\pm$16 GCs with a specific frequency of 4.2$\pm$0.8 and a specific mass of 0.23$\pm$0.01, which is relatively high compared to other galaxies of similar mass and environmental conditions. The spatial and azimuthal distributions of subpopulations reveal strong evidence of previous interactions within the host galaxy. The color distribution of GCS in NGC4262 shows a gradient of -0.05$\pm$0.01 within 5.5\arcmin, supporting the notion of past interactions and evolutionary transitions. PRG NGC4262 conforms to the overall trend of the GCS mass with respect to the halo mass. Furthermore, our investigation of the global scaling relations between GCS and host galaxy parameters provides further support for the hypothesis that PRGs are an intermediate phase connecting ongoing mergers and quiescent galaxies.

\end{abstract}


\section{Introduction}

\begin{figure*}
    \centering
    \includegraphics[width=2\columnwidth]{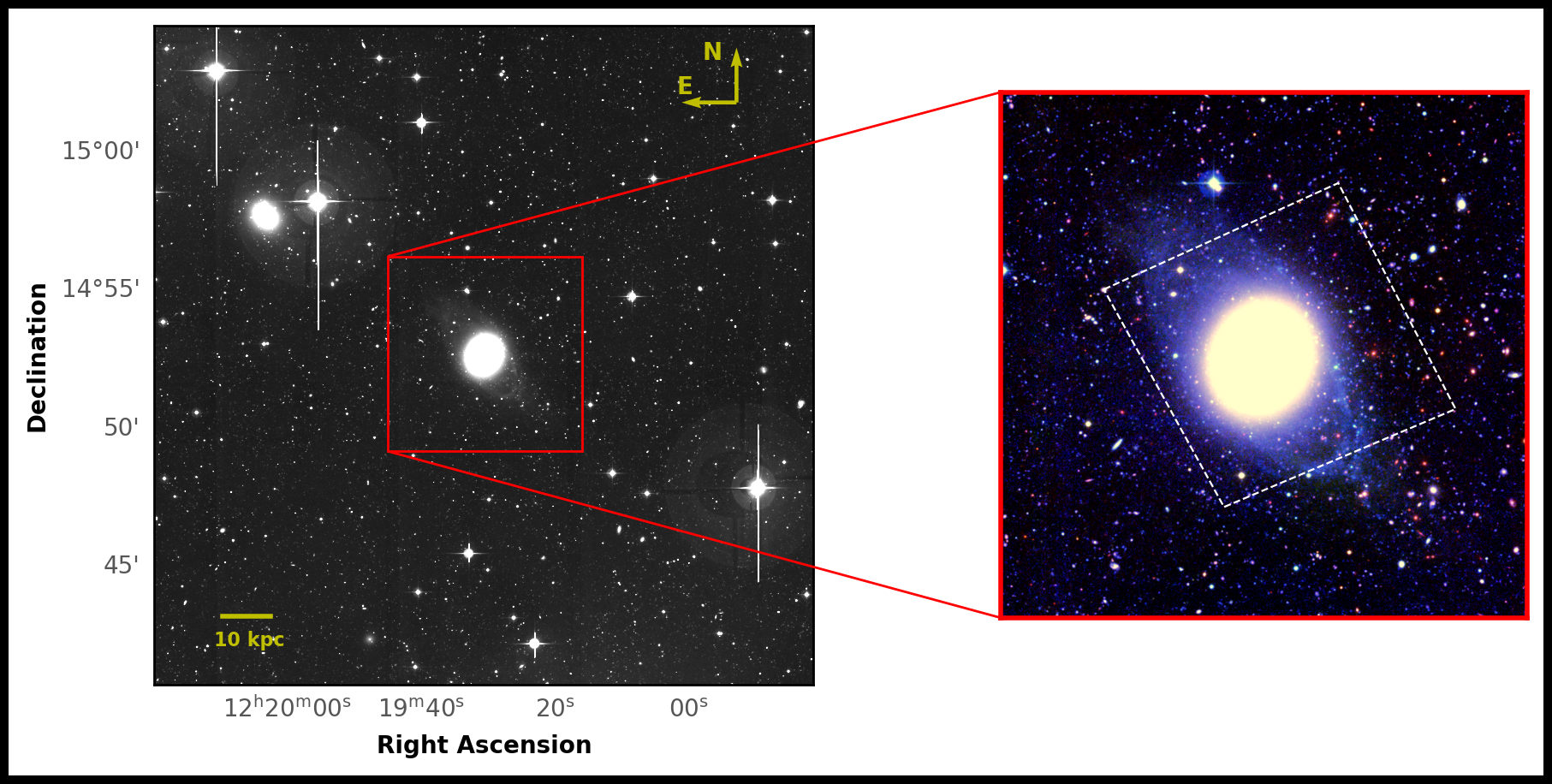}
    \caption{The optical $\it g$-band image of NGC4262 using the CFHT. A horizontal yellow solid line on the bottom left corner indicates the image scale of 10 kpc. The inset shows the optical color image with a field of view of 7\arcmin x 7\arcmin.  Blue, green, and red colors represent the CFHT  {\it u, g } and {\it i} bands, respectively. This image of the galaxy reveals an optically faint ring component that remains undetected in the earlier optical images. In the inset image, the white dotted line represents the HST ACS observed region.}
    \label{fig:igu_color}
\end{figure*}

NGC 4262 is a lenticular type (S0) galaxy exhibiting peculiar features located in the Virgo cluster region with a redshift of 0.0045 \citep{whitmore1990}. The exploration of the gas-rich lenticular galaxy NGC 4262 in the Virgo cluster region began with \citet{1983Giovanardi}. Subsequent studies by \citet{1985krum} and \citet{1987vanDriel} delved further into the galaxy's properties and detected a noticeable amount of atomic hydrogen surrounding NGC 4262. From these observations, \citet{whitmore1990} proposed that NGC 4262 could potentially be classified as a kinematically possible Polar Ring Galaxy (PRG). \par
PRGs mostly belong to the lenticular type, having a ring of gas and stars orbiting in a nearly polar/orthogonal plane. They are composed of two distinct components -- a host without gas \citep{finkelman2012} and a polar ring or a disk consisting of gas, dust, and stars \citep{reshkinov1997,2019Reshetnikov}. The major formation scenarios proposed for the PRGs are major galaxy merging \citep{1998ApJ...499..635B,2003Bournaudcombes}, tidal accretion of mass from a nearby companion \citep{1997A&A...325..933R,2003Bournaudcombes}, and cold gas accretion from the intergalactic medium \citep{2008ApJ...689..678B,2019Egorov}. Numerous studies, including (eg: \citet{reshkinov1997,2014Iodice,2014Khoperskov,2015Moiseev,2015Spavone,ordernes2016A&A...585A.156O,2020Smirnova,2023Deg,2024Smirnov}) have broadly examined kinematics, structure, dark matter content, halo shape, star formation, and other various aspects of PRGs. Even though extensive studies have been conducted on the PRGs, the study of their Globular Clusters (GCs) still remains unexplored. \par 
GCs are the oldest stellar systems in a galaxy and hence referred to as ''fossils". Thus, by exploring the GC distribution in a galaxy, we better understand the host galaxy properties. \citep{forbed1997,2brodieandstrader006ARA&A..44..193B,Beasley2020rfma.book..245B}. GCs have an average age greater than 10 Gyr and are luminous point-like objects that can be detected even at distances of hundreds of megaparsecs \citep{2011Peng,2014Harris}. The globular cluster system (GCS) encompasses all the GCs situated in the spheroid and halo regions of a galaxy \citep{Harris1991,ashmanzepf1992}. In the early-type galaxies (ETGs), the color distribution of GCs often shows bimodality, which indicates the presence of two subpopulations \citep{Harris1991,Peng2006,Hagrisrhode2014,2022deBritosilvajplus}. Multi-phase collapse, accretion, and merging of gas-rich galaxies are the proposed theories for the origin of this bimodality \citep{forbed1997,2008Bournaud,2014LiandGnedin,2022LiandVogelsberger}. Also, few studies have investigated the presence of multiple subpopulations due to different galaxy interactions \citep{2015Caso,2012Blom,2020Escudero,2022Escudero}.
The properties of GCS, including color distribution, specific frequency, radial distributions, and azimuthal distributions, are best studied with different Hubble-type galaxies \citep[e.g.][]{2004Rhodeandzepf,2011Hargis,2011forbe,2012Blom,sreeja2014,2018Sesto,Beasley2020rfma.book..245B,2022Buzzosplus,2023harris,2024Caso}. Strong correlations between the characteristics of the GCs and the host haloes have been broadly studied in the last few decades \citep[references therein]{2008peng,2015Harris_hudson,2017Forbes,2022Reina}. The confirmation of GCs at distances over 13 times the effective radius of their host galaxy makes them effective probes for the structure of dark matter haloes \citep{2016Alabi}. A detailed examination of GC properties can provide vital information to galaxy evolution scenarios and is a powerful tool for tracing the structure and history of galaxies. \citep{forbed1997,2brodieandstrader006ARA&A..44..193B,2019Kruijssen,Beasley2020rfma.book..245B}. \par

Conducting a thorough study on the GCS of the nearest PRG, NGC 4262, presents an excellent opportunity to investigate the GCS and its properties associated with the PRGs. In this work, we made use of the optical data in {\it u, g, i, and z } of the Canada-France-Hawaii Telescope (CFHT) to explore the host galaxy and its GCS of NGC 4262. Table \ref{tab: Table:1} lists the basic characteristics of NGC 4262. Section \ref{sec:data} summarizes the CFHT observations and the data reduction procedure. Section \ref{sec:phot_and_GCS} presents the surface brightness profiles of NGC 4262 and the old stellar population analysis using the CFHT data. Finally, sections \ref{sec:discussion} and \ref{sec:summary} give the discussion and summary of this work.
A flat Universe cosmology is adopted throughout this paper with $H_{0}$ = 71 $kms^{−1}$ $Mpc^{−1}$ and $\Omega_M$ = 0.27 \citep{2011Komatsu}.

\begin{table}
\caption{Physical parameters of galaxy NGC 4262. References are 1. \citet{2008sdssdr6}, 2. \citet{1991morph_and_redshift}, 3. \citet{2018Haynes}, 4. \citet{cappellari_atlas3d2011}, 5. \citet{2013Harris}, 6. \citet{2007vandenBergh}, 7. \citet{2015salo}}
\scalebox{1}{
\begin{tabular}{ll}
\hline
\textbf{Parameter}                  & \textbf{Value}   \\ \hline
\multicolumn{1}{|l|}{RA$^{1}$}       & \multicolumn{1}{l|}{12h19m30.575s}  \\ 
\multicolumn{1}{|l|}{DEC$^{1}$}           & \multicolumn{1}{l|}{ +14d52m39.56s}  \\ 
\multicolumn{1}{|l|}{Type$^{2}$}          & \multicolumn{1}{l|}{SB0}  \\ 
\multicolumn{1}{|l|}{Radial velocity$^{3}$}& \multicolumn{1}{l|}{ 1367 ± 2 kms$^{-1}$} \\ 
\multicolumn{1}{|l|}{Distance$^{4}$}       & \multicolumn{1}{l|}{15.5 Mpc}  \\ 
\multicolumn{1}{|l|}{$M_{V}^{5}$}   & \multicolumn{1}{l|}{-19.51 mag}  \\
\multicolumn{1}{|l|}{$M_{B}^{6}$}   & \multicolumn{1}{l|}{-18.59 mag}  \\

\multicolumn{1}{|l|}{Apparent diameter$^{2}$}& \multicolumn{1}{l|}{11 kpc}  \\ 
\multicolumn{1}{|l|}{Position angle$^{7}$}   & \multicolumn{1}{l|}{158$^{\circ}$}  \\
\multicolumn{1}{|l|}{Metric scale}   & \multicolumn{1}{l|}{75 pc/"}  \\

\hline
\end{tabular}}

\label{tab: Table:1}
\end{table}

\section{Observations and data reduction}
\label{sec:data}
 
\begin{table}
\caption{Log of CFHT observations.}
\label{tab:Log_table}
\scalebox{1}{
\begin{tabular}{|l|l|l|l|}
\hline
\textbf{Filter} & \textbf{Obs. date} & \textbf{Seeing (")} &  \textbf{Exp. time (s)}  \\ \hline
{\it u}               & 2011-04-01         & 0.88              & 6402 \\ 
{\it g}               & 2010-04-13         & 0.87               & 3170 \\ 
{\it i}               & 2010-05-17         & 0.6                & 2055 \\ 
{\it z}               & 2010-03-15         & 0.89               & 4400 \\ \hline
\end{tabular}}
\end{table}
NGC 4262 was observed as part of the Next Generation Virgo Cluster Survey \citep[NGVS]{Ferrarese2012}. Deep optical images in {\it u, g, i} and {\it z } bands are obtained from the NGVS, taken with the MegaCam instrument. The camera comprises 40 CCDs with a plate scale of 0.187"/pixel. The MegaCam images have been reduced and analyzed using the Elixir-LSB pipeline \citep{Ferrarese2012}. The pipeline carries out an astrometric and photometric calibration for the MegaCam images. The available MegaCam images were utilized for background subtraction in order to create a comprehensive background map. This map was then scaled to match the corresponding sky level and subtracted from the NGVS images. The background-subtracted images are combined using the artificial skepticism combined method. The NGVS images are calibrated to a photometric zero point of 30 mag, such that AB magnitudes are given by:
\begin{equation}
    m(AB) = -2.5 \times \log_{10}(DN) + 30.0,
\end{equation}

where DN is the number of counts measured in the image \citep{Ferrarese2012}. A detailed description of the NGVS, data procurement, reduction pipeline employed, and processing strategies can be found in \citet{Ferrarese2012,durell_ngvs2014} and \citet{cantiello_ngvs2018}. The observation log is tabulated in Table \ref{tab:Log_table}, and in Figure \ref{fig:igu_color}, the inset represents the optical color composite image of NGC 4262. \par

\citet{2009Jordan} presented catalogs of GC candidates for 100 ETGs using the HST/ACS survey program of the Virgo cluster. From the catalog, we obtained the photometric catalog of GC candidates of NGC 4262 in the filters g (F475W) and z (F850LP). Note that the ACS field of view covers only a smaller area ($\sim $3\arcmin) than the CFHT field of view for the galaxy NGC 4262. To account for the galactic extinction in optical data, we used the \citet{1989Cardelli} extinction law with the  $A_v$ value of 0.098 \citep{2011Schalfyfinkbeinr}. 

\section{Photometric analysis of NGC 4262 and its GCS}
\label{sec:phot_and_GCS}
\subsection{Galaxy surface photometry}
\label{sec:surface_PHOTOMETRY}
The morphology of NGC 4262 is an early-type S0 with Hubble type T = -2.7 \citep{cappellari_atlas3d2011}. \citet{Ferrarese2006} mentioned the presence of a dominant stellar bar, a small regular dust disk, and thin dust filaments in NGC 4262 using the HST data. According to \citet{Buson2011}, the galaxy is surrounded by a UV ring structure and is not visible at optical wavelengths. However, the RGB optical image presented in this study (Figure \ref{fig:igu_color})  clearly reveals the presence of the ring structure mentioned. Although NGC 4262 has a multifaceted isophotal structure, examining its one-dimensional light profile can be insightful.
\begin{figure}
    \centering
    \includegraphics[width=0.65\columnwidth]{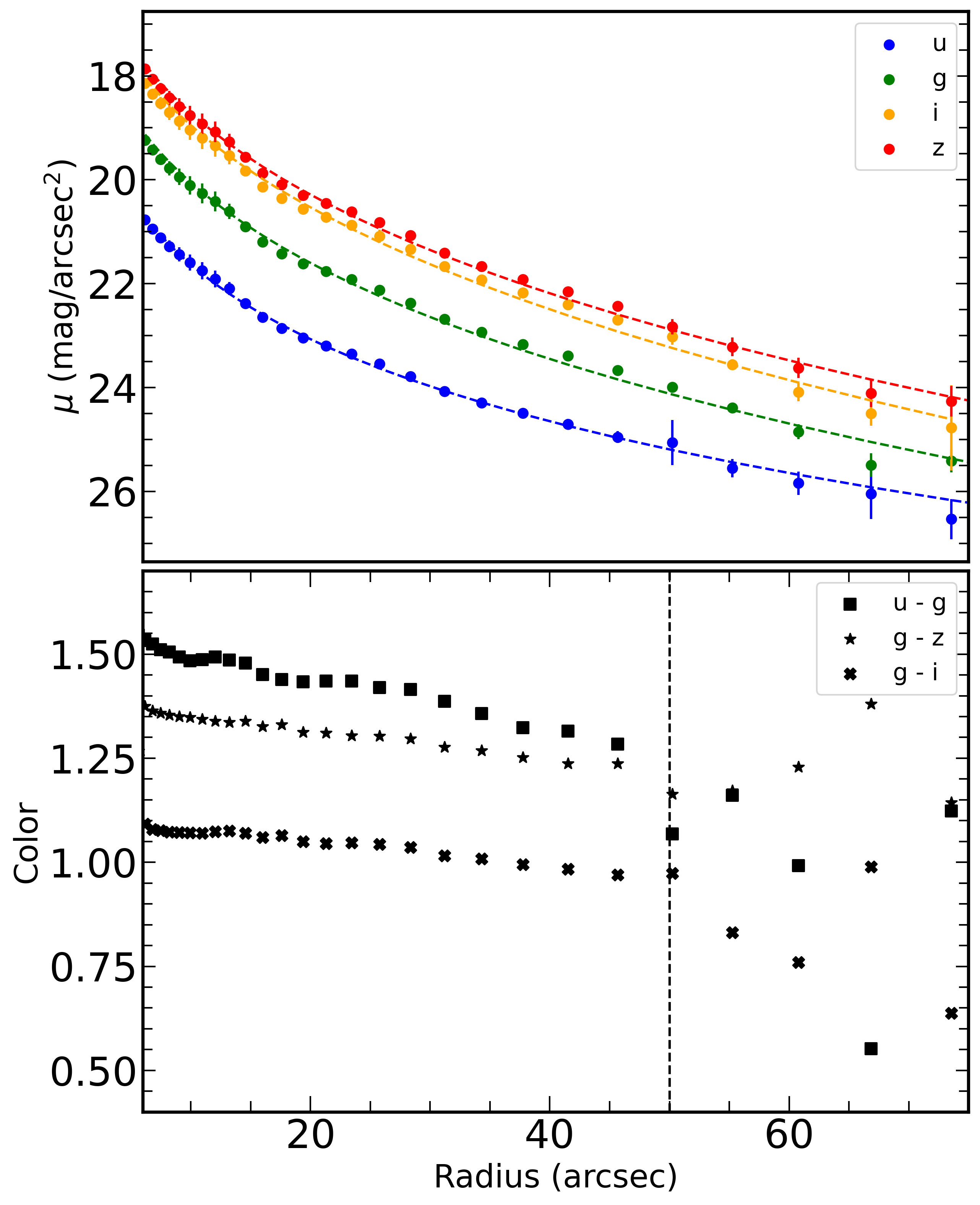}
    \caption{Surface-brightness profiles of the NGC 4262 obtained in {\it u, g, i}, and {\it z} -bands are shown in the upper panel. The dashed lines represent Sérsic models fitted to the galaxy profile. The lower panel represents the optical color profiles  {\it u - g, g - z } and {\it g - i} as a function of radius. We can observe a significant variation in the trend of each color profile at a 50" radius (vertical dotted line), which may be connected with the ring structure.}
    \label{fig:sbp1}
\end{figure}

\begin{figure}
    \centering
    \includegraphics[width=0.65\columnwidth]{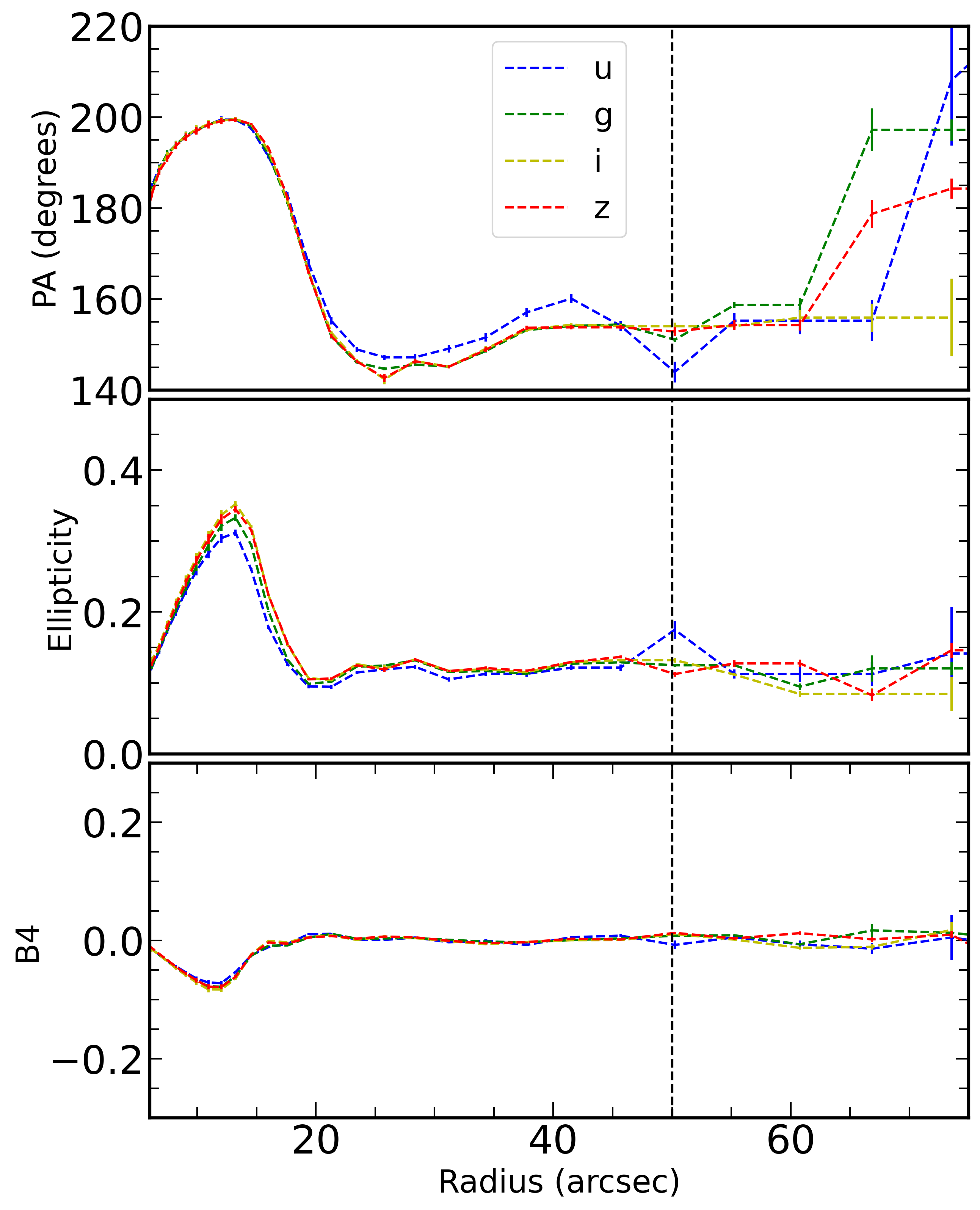}
    \caption{Upper, middle, and lower panels show the variations of the isophotal parameters PA,  $\epsilon$, and B4 in  {\it u, g, i}, and {\it z} -filters as a function of radius, respectively. The radius of 50" is indicated by a vertical dotted line.}
    \label{fig:sbp2}
\end{figure}
From the isophotal analysis, we obtained the surface brightness and color profiles of NGC 4262 in the optical {\it u, g, i} and {\it z} bands along the major axis. The light from the galaxy was modeled using Python \citep[Elliptical Isophote analysis]{jedrzejwski1987} in all the filters. The surface brightness profiles in {\it u, g, i} and {\it z} bands of the galaxy are shown in Figure \ref{fig:sbp1} upper panel.  The surface brightness profile is fitted with a Sérsic function \citep{1963Sersic} and is given below,
\begin{center}
\begin{equation}
\mu(r) = \mu_{e} +\left(\frac{2.5 b_{n}}{ln 10}\right)\left[\left(\frac{r}{r_e} \right)^{(1/n)} - 1\right] 
\end{equation}
\end{center}
where $r_{e}$ is the effective radius of the galaxy, $\mu_{e}$ is the surface brightness at the effective radius, n is the Sérsic index, and $b_{n}$ = 1.9992n − 0.3271. Our findings reveal that NGC 4262 has an average optical effective radius of approximately 11.8 $\pm$  0.4" (0.9 $\pm$ 0.03 kpc). \citet{2013Harris} estimated $r_{e}$ for the galaxy using HST optical images and is 0.82kpc. The parameters for the best-fit result are listed in Table \ref{tab:sbtab}. In the lower panel of Figure \ref{fig:sbp1}, the colors  {\it u - g, g - z } and {\it g - i} as a function of radius are plotted. At a radius of 50"  (3.8 kpc), we noticed a significant variation in the trend of each color profile. This variation in the optical color profiles indicates the presence of a ring structure. \par
Figure \ref{fig:sbp2} displays the isophotal parameters, including position angle (PA), Ellipticity ($\epsilon$), and the Fourier amplitudes of the deviation from perfect ellipses (B4). Within a $\sim $20" (1.5 kpc) radius, notable variations were observed in the parameters related to the rise and fall of PA, $\epsilon$, and changes in B4. These variations suggest the existence of a dominant stellar bar, as mentioned in \citet{Ferrarese2006}. The isophotal parameters remain at a constant value between 20" and 50" radii.  Beyond $\sim$50", minor variations in the isophotal parameters were also observed. The color and isophotal parameters show significant variations after a radius of 50", which is also visually evident in the RGB image (Figure \ref{fig:igu_color}). The average {\it g - i} color within the 50" radius is estimated to be 1.00 $\pm$ 0.01, whereas in the 50" to 75" range, it measures 0.72 $\pm$ 0.02. This indicates that the outer structure (ring) is bluer than the central component (host). The observed color difference aligns with the results of earlier studies conducted by \citet{2015ReshetnikovCombes2014} and \citet{2024akr}, which indicated that ring exhibit color values similar to spiral galaxies, while the host displays color values of ETGs.

In addition to these results, we obtained the position angle and ellipticity of host galaxy NGC 4262 as  PA = 150 $\pm$  0.7$^{\circ}$, and $\epsilon$ = 0.13 $\pm$ 0.003. The obtained values closely resemble those reported by \citet{2015salo}, which are PA = 158$^{\circ}$ and $\epsilon$ = 0.09.

\begin{table}
\caption{Presents the results obtained from fitting the surface brightness profiles of NGC 4262 with Sérsic profiles in the {\it u, g, i}, and {\it z} filters.}
\scalebox{1}{
\begin{tabular}{|l|l|l|l|l}
\cline{1-4}
filter & $\mu_{e} (mag/$"$^{2}) $& $R_{e}($"$)$ & n   &  \\ \cline{1-4}
{\it u}      & 22.6 $\pm$ 0.1                    & 16.4 $\pm$ 1.1           & 7.3 $\pm$ 1.1 &  \\
{\it g}      & 20.2 $\pm$ 0.1                   & 10.6 $\pm$ 0.7           & 3.9 $\pm$ 0.4 &  \\
{\it i}      & 18.9 $\pm$ 0.1                    & 10.1 $\pm$ 0.6           & 3.1 $\pm$ 0.3 &  \\ 
{\it z}      & 18.7$\pm$ 0.1                    & 10.1 $\pm$ 0.5           & 3.6 $\pm$ 0.3 &  \\ \cline{1-4}
\end{tabular}}
\label{tab:sbtab}
\end{table}

\subsection{The Globular Cluster System}
\label{sec:old_stellar_population_analysis}
\subsubsection{Source detection and photometry}
\label{sec:dectection_photometry}
In this section, we explore the GCS of NGC 4262 using the CFHT wide-field optical images. At a distance of Virgo Cluster, i.e., 16.5 Mpc \citep{2007Mei}, most GCs are expected to seem like point sources in MegaCam images \citep{durell_ngvs2014,2019MNRAS.483.1470R}. We used SExtractor software \citep{Bertin_arnouts1996} for source detection in {\it u, g, i} and {\it z} band images. To improve the source extraction procedure, we modeled the galaxy light of NGC 4262 and subtracted it from the corresponding mosaic image using the elliptical isophotal analysis, as mentioned in section \ref{sec:surface_PHOTOMETRY}. We detected the point sources with a threshold of 3$\sigma$ in each filter. The sources detected in all four filters are considered as the initial sample. We used the criteria in SExtractor named CLASS/GALAXY parameter (> 0.5) to remove the background objects. Approximately 50$\%$ of the detections were subsequently removed based on the criteria. To remove spurious detections such as image artifacts and statistical events, we crossmatched individual sources by their positions between the {\it u, g, i,} and {\it z} images. Aperture photometry, including local sky background subtraction, was performed on the identified sources on the optical images with galaxy light, using an 8-pixel aperture diameter (1.48", twice the seeing). The aperture correction was applied using the curve of growth of unsaturated stars in the field \citep{Ferrarese2012,durell_ngvs2014,2015Liu,2019ko}. Based on the analysis conducted above, a total of 5600 objects were obtained for further analysis.
 
\subsubsection{Selection of GCs}
The GC selection for NGC 4262 is based on the magnitude and color of individual detected sources. To identify the point-like sources from the catalog, we used the Concentration index \citep[C,][]{2011Peng}. The magnitude difference at the 4 and 8-pixel diameter (0.74" and 1.48") aperture was used to select the point-like sources. For all point-like objects, C is expected to be consistent with zero after applying the aperture corrections. In this work, we assumed a scatter of 0.1 mag to comply with the errors in the photometry of the sources \citep{durell_ngvs2014,2019ko}. \par
\begin{figure}
    \centering
    \includegraphics[width=\columnwidth]{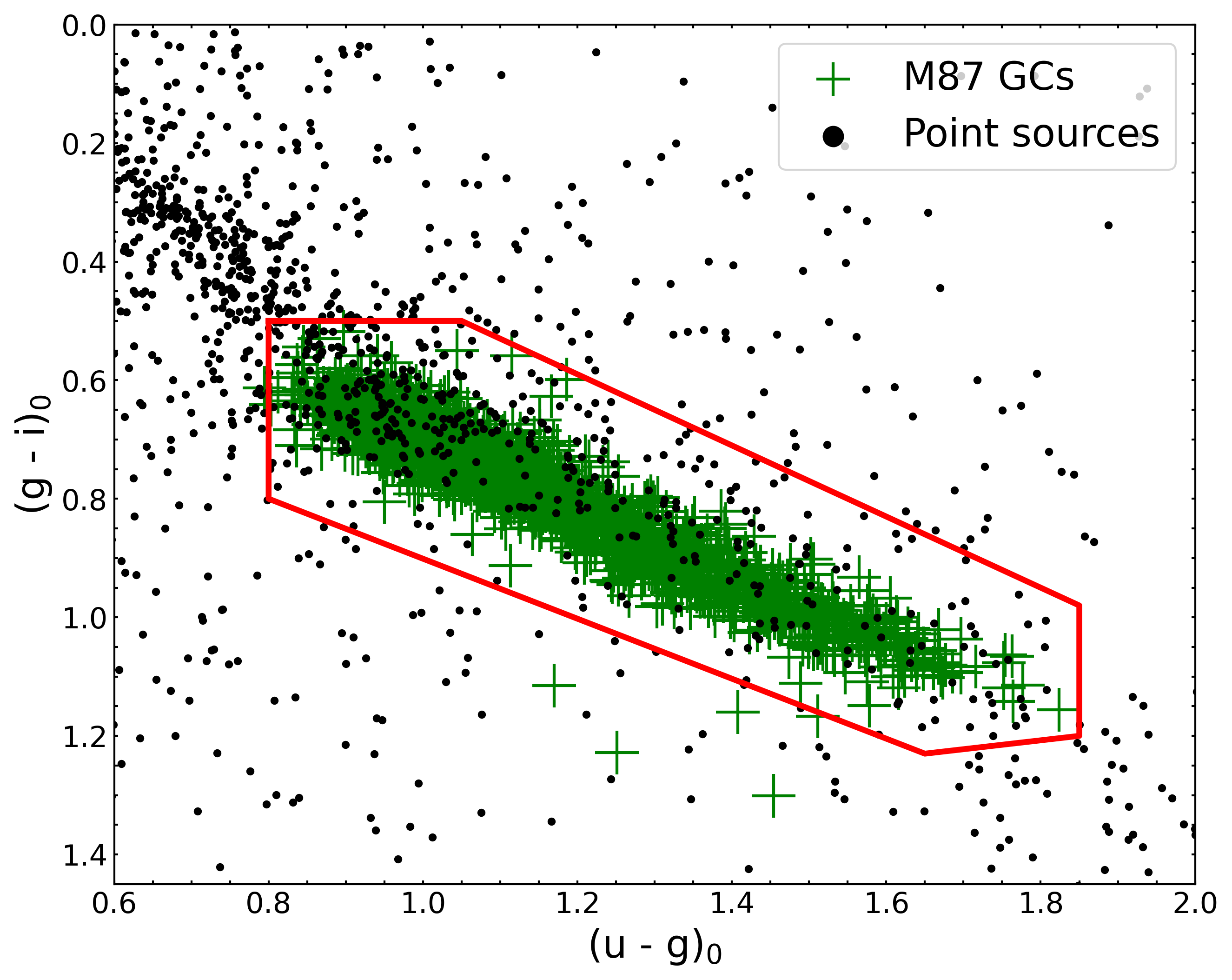}
    \caption{(u-g)$_{0}$ - (g-i)$_{0}$ CCDm for the sources with −0.10 < C < 0.10 in the galaxy field are shown in black-filled circles. The red-colored polygon indicates the selection criteria for the GCs obtained from \citet{2017Lim}. The GCs of M87 obtained from \citep{2016Powalka} are overplotted with green plus symbols.}
    \label{fig:ccd}
\end{figure}

Later, we used the proper motions of the objects from the {\it Gaia} Data Release 3 \citep[{\it Gaia} DR3,][]{2022Ygaiadr3} to remove definite stars from our data sample \citep{2020Voggel,2022Buzzosplus}. Through this approach, we identified 67 foreground stars and removed from the sample. After identifying the initial sample of point-like sources without any Galactic stars, we applied the galactic extinction corrections obtained from \citet{2011Schalfyfinkbeinr} to the calibrated magnitudes. 
To obtain sample objects with low color errors, we considered a photometric quality cut to the magnitude. The magnitude errors obtained at  i$_{mag}$ = 23.5 are lower than 0.1 mag. In the sample, we applied color limits to select the GC candidates of galaxy NGC4262. \citet{2017Lim} presented a guideline for GC selection based on {\it u g i} photometry using spectroscopically confirmed GCs in M87 \citep{2011Strader,2015Zhang}. Figure \ref{fig:ccd} shows the color-color diagram (CCDm), and the red lines represent the obtained limits from \citet{2017Lim}. In this study, we adopted the same selection criteria for the GC candidates of NGC 4262. To provide a reference, we included the confirmed GCs in M87 \citep{2016Powalka} by overplotting them in the CCDm. All magnitudes are extinction-corrected and represented with a superscript of zero. We obtained 325 GC candidates as our final sample from all the above analyses.

\subsubsection{One--dimensional radial distribution of GCS}
\label{sec:radialprofile}
\begin{figure}
    \centering
    \includegraphics[width=\columnwidth]{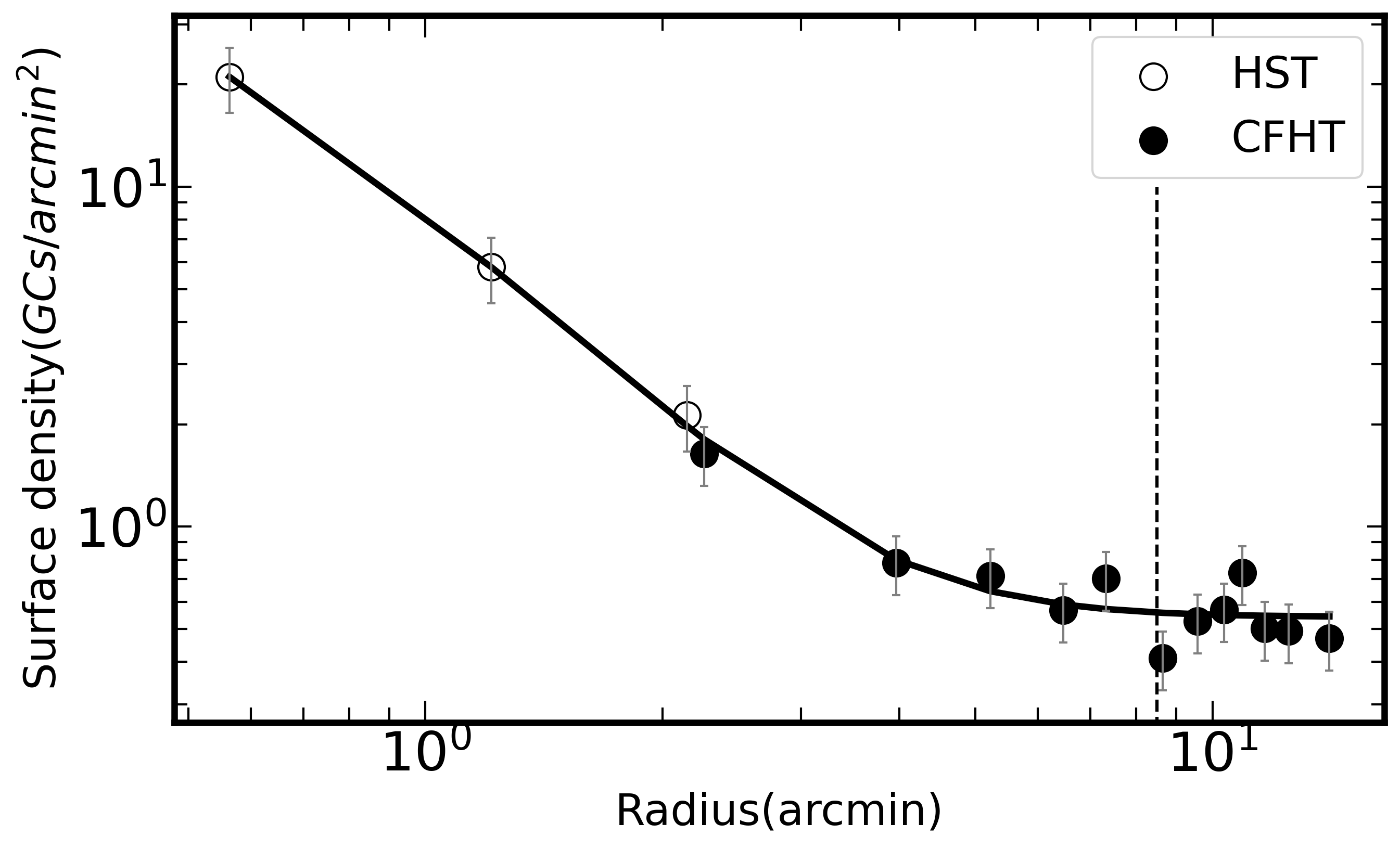}
    \caption{Surface density profile for the GCS of NGC 4262. The plot incorporates HST (open circles) and CFHT (filled circles) data. A Sérsic profile is fitted and is shown with a solid line. The best-fit parameters are listed in Table \ref{tab:sbp_subpop}. The surface density of the GCS reaches the background level at 8.5 $\pm$ 1\arcmin, indicated by a dotted vertical line.}
    \label{fig:sdp}
\end{figure}

The surface density profiles demonstrate the spatial distribution of GCs in the galaxy. Since there is a saturation in the galaxy's center by the host galaxy light in CFHT observations, we utilized GCs from HST observations \citep{2009Jordan}. We generated the surface density profile for NGC 4262 by combining GCs from the HST for the central region, covering a radius of approximately 3$\arcmin$, and those from CFHT for the outer regions, extending up to 16$\arcmin$ from the galaxy's center. The total GCs obtained were radially binned with the effective covered area and obtained the density profile as discussed in \citet{sreeja2014,2016Kartha}.
We observed a dwarf galaxy, IC 781, in the field at a distance of 12.2$\arcmin$ from the center of NGC 4262. We also found that IC 781 has a distance of 24.82 +/- 1.77 Mpc \citep{2014Toloba}. Using the relationship of galaxy effective radius and spatial GC extent from \citet{sreeja2014}, we have roughly estimated the GC extent of IC 781 as 2$\arcmin$. Then, we removed the corresponding GCs around IC 781 inside 2$\arcmin$  from the main sample and accounted for the respective area correction. Poisson statistics gives the errors associated with the surface density distribution. The GC surface density distribution is fitted using a combination of a Sérsic profile and a background parameter. The profile used for the analysis is given below,
\begin{equation}
    N(R) = N_e exp[-b_n(R/R_e)^{n}-1] + Bg 
\label{eq:1}
\end{equation}
where $N_{e}$ is the surface density of the GCs at the effective radius $R_{e}$, n is the Sérsic index or the shape parameter for the profile, $b_{n}$ is given by the term 1.9992n − 0.3271, and Bg represents the background value. Figure \ref{fig:sdp} displays the surface density profile for the GCS of NGC 4262 using the HST and CFHT data. The GC surface density is fitted with the Sérsic profile (equation \ref{eq:1}). The best-fit parameters are listed in Table \ref{tab:sbp_subpop}. We obtained that, at a radius of 8.5 $\pm$  1\arcmin, the GCS surface density flattens to a constant background value of 0.45 $\pm$ 0.04 objects per $arcmin^{2}$. Hence, the GCS extent of NGC 4262 obtained from its radial density profile is 8.5 $\pm$ 1$\arcmin$ (38 $\pm$ 5 kpc).  From the total of 325 GCs obtained from the CFHT data analysis, we identified that 112 GCs lie inside the estimated extent. The GCs outside the extent are considered as background population.

\subsubsection{Color distribution and subpopulations}
\begin{figure}
\centering
\includegraphics[width=0.9\columnwidth]{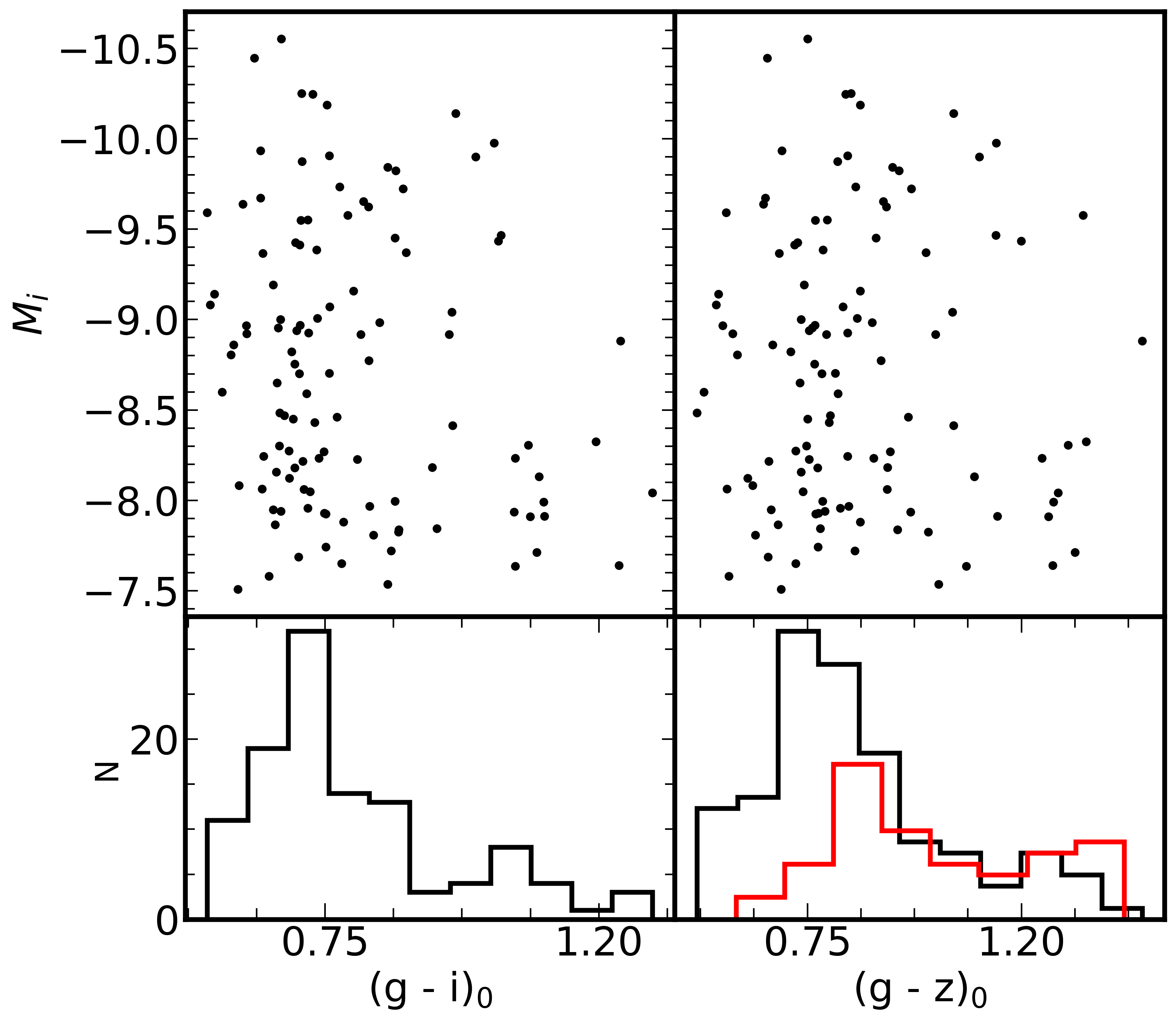}
\caption{ The CMDs for the selected GC candidates within the estimated GCS extent using CFHT data. The bottom panel represents the corresponding $(g - i)_0$ and $(g - z)_0$  histograms of the GC candidates. The red dotted line represents the HST data obtained from \citet{2009Jordan}.}
\label{fig:hist}
\end{figure}

\begin{figure}
    \centering
    \includegraphics[width=\columnwidth]{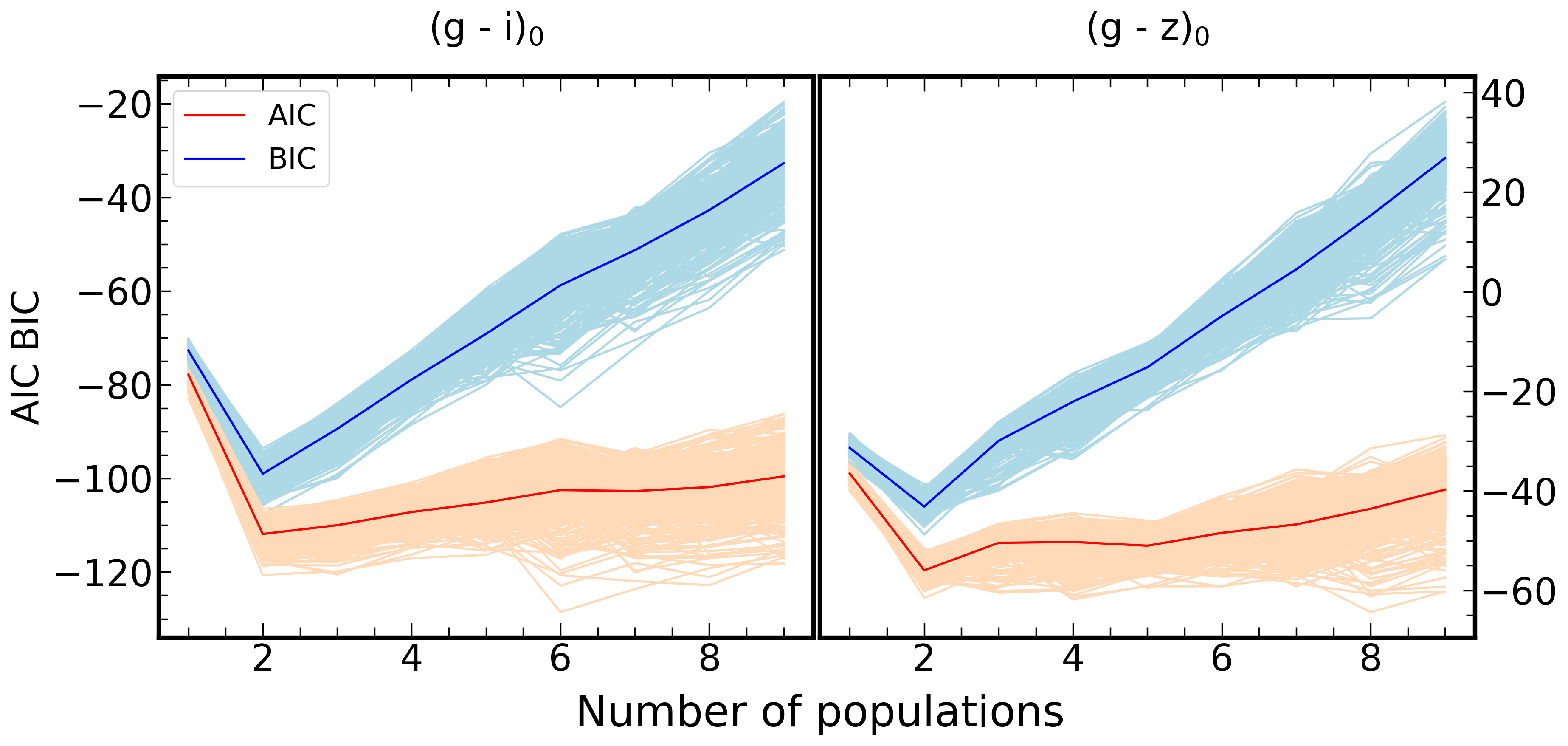}
    \caption{Shows the AIC and BIC values for 500 experiments for (g - i)$_{0}$ and (g - z)$_{0}$ color data in the left and right panels, respectively. The bold blue and red lines represent the mean values obtained from the 500 experiments of BIC and AIC, respectively.}
    \label{fig:aicbic}
\end{figure}

From the surface density profile, we have estimated that the GCS extent of galaxy NGC 4262 is 38 $\pm$ 5 kpc. Here, the selected 112 GCs are defined as the final sample GCs of NGC 4262 from CFHT. The Color Magnitude Diagrams (CMDs) and color distributions of the selected GC candidates within the measured GCS extent are shown in Figure \ref{fig:hist}. In the Figure, we added HST GCS in the (g - z)$_{0}$ color histogram for comparison. We observed clear separation in the red and blue subpopulations in the color distributions. To evaluate the multiple populations in the color distribution, we used different statistical models. The number of components or subpopulations in our dataset is determined by two probabilistic statistical models, the Akaike and Bayesian Information Criterion \citep[AIC and BIC]{1974Akaike_AIC,1978Schwarz_BIC}. In order to correct the background population in the color distributions, we explored the same approach as discussed in \citet{2022Escudero}. We generated the color distribution of the background population in an annular region outside the estimated GCS extent. In the (g - i)$_{0}$ and (g - z)$_{0}$ color distributions of the background populations, we considered bin sizes of 0.05 mag and identified the number of objects in each bin. Then, in the GC populations of the same bin, objects were randomly removed according to the background populations. We conducted 500 experiments to analyze the AIC and BIC models to identify multiple populations. The results of the AIC BIC model analysis in both (g - i)$_{0}$ and (g - z)$_{0}$ colors shown in Figure \ref{fig:aicbic} clearly indicate bimodality in the system.\par

\begin{figure}
    \centering
    \includegraphics[width=\columnwidth]{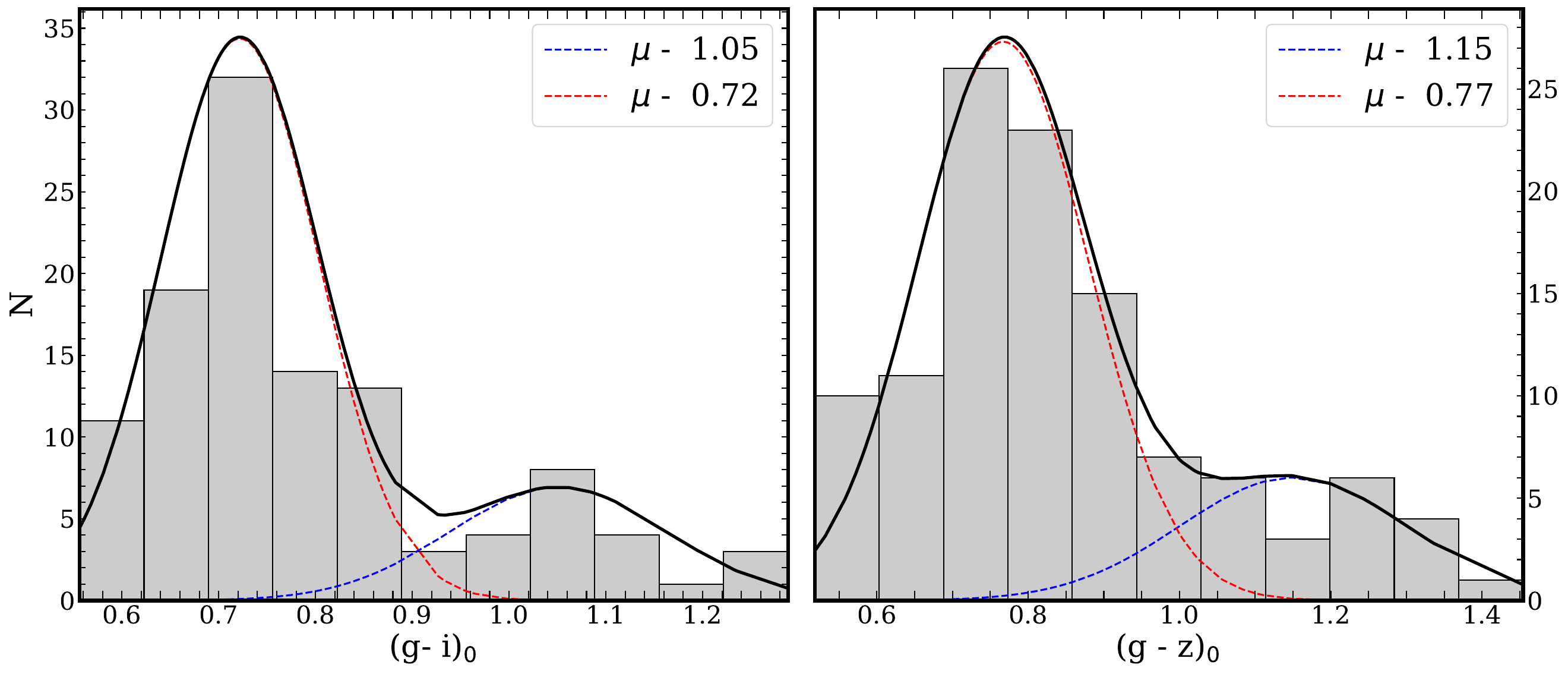}
    \caption{Color histogram of GCs modeled with GMM. Blue and red dashed lines indicate the Gaussian components fitted by sklearn mixture package, where their sum is indicated by a black solid line. The mean values are labeled inside the Figure.}
    \label{fig:gmmfit}
\end{figure}

In addition, we used the Gaussian Mixture Modelling \citep[GMM]{2010Muratov_GMM} algorithm to quantify the color distributions. GMM is a probabilistic model in which all data points are generated from a mixture of finite Gaussian distributions with parameter estimates. We fit GMM to the color distributions using the expectation-maximization algorithm through the open-source machine learning library SCIKIT-LEARN for python \citep{2012Pedregosa_scikit}. Using the GMM algorithm by considering a bimodal color distribution in our GCS, we identified the peaks of blue and red subpopulations in both (g - i)$_{0}$ and (g - z)$_{0}$ colors. Figure \ref{fig:gmmfit} represents the color histograms for the two color distributions. The (g - i)$_{0}$ and (g - z)$_{0}$ color values exhibit distinct peaks, with the red subpopulation peaking at 1.05 and 1.22, while the blue subpopulation peaks at 0.72 and 0.79, respectively. We compared these peak values of the distribution with the expected average color values for the blue and red subpopulations of GCS based on the relations presented by \citet{Peng2006}   and \citet{2011Faifer}.
The estimated color peaks in this study are closely aligned with the color values obtained from \citet{Peng2006} and \citet{2011Faifer} ((g - i)$_{0}$ mean values for the blue and red subpopulations are 0.75 $\pm$ 0.29 and 1.0 $\pm$ 0.4, while the (g - z)$_{0}$ mean values are 0.91 $\pm$ 0.08 and 1.30 $\pm$ 0.11, respectively.) within the margin of error. As a result, we found that the blue subpopulation accounted for 64 $\pm$  8 percent, and the red subpopulation accounted for 36 $\pm$  7 percent of the total data.

\subsubsection{Spatial and azimuthal distribution of subpopulations}

\begin{figure}
    \centering
    \includegraphics[width=\columnwidth]{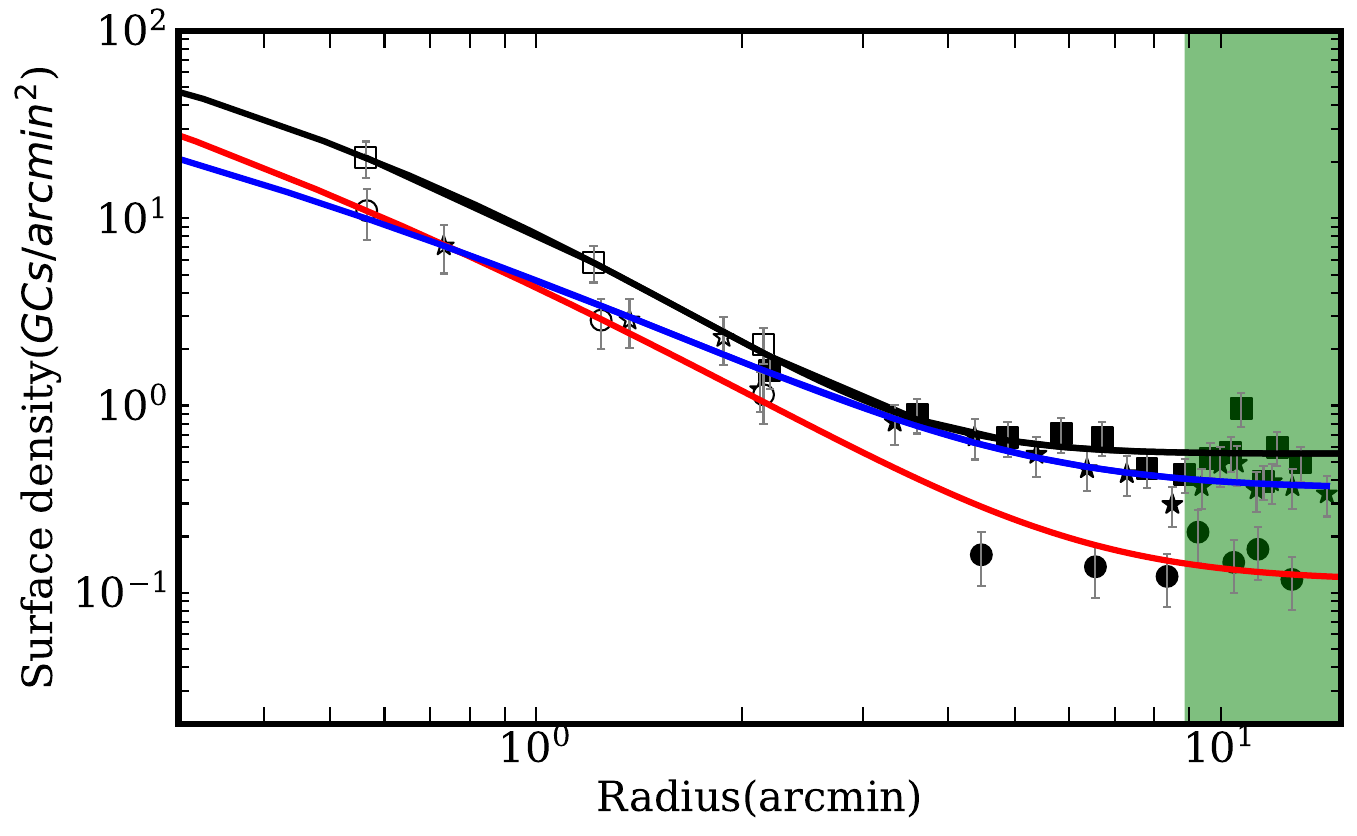}
    \caption{The radial density profile for the GC subpopulations of NGC 4262 is shown here.  The square, star, and circle markers correspond to the total GCs, blue, and red subpopulations, respectively. The fitted Sérsic profiles for the total GCs, blue, and red subpopulations are represented by black, blue, and red solid lines, respectively. The CFHT and HST data points are plotted with filled and open markers, respectively.}
    \label{fig:sdp_sub}
\end{figure}

In section \ref{sec:radialprofile}, we analyzed and studied the radial density distribution of total GCS in NGC 4262. Here, we use the same method to examine the radial density distribution of GC subpopulations. Figure \ref{fig:sdp_sub} shows the radial surface density profiles for blue, red, and total GCs with the best-fit surface density model. The estimated results are given in Table \ref{tab:sbp_subpop}. From our analysis, we found that the red subpopulation is more prominent than the blue subpopulation in the galactic center, as shown in Figure \ref{fig:sdp_sub}. However, as the distance from the galaxy center increases, the blue subpopulation begins to dominate over the red subpopulation. \par

\begin{table}
\caption{Results obtained from the Sérsic profile fitting in the total GCs and its subpopulations of NGC 4262.}
\begin{tabular}{|l|l|l|l|l}
\cline{1-4}
Population & $R_{e} (\arcmin) $ & $n$ & Bg   &  \\ \cline{1-4}
Total     & 0.97 $\pm$ 0.06       & 2.7 $\pm$ 0.5         & 0.5 $\pm$ 0.04 &  \\
Blue      & 1.8 $\pm$ 0.8         & 2.3 $\pm$ 1.7        & 0.3 $\pm$ 0.09 &  \\
Red       & 1.7 $\pm$ 1.1         & 5.8 $\pm$ 4.5         & 0.1 $\pm$ 0.05 &  \\ 
\cline{1-4}
\end{tabular}
\label{tab:sbp_subpop}
\end{table}

\begin{figure}
    \centering
    \includegraphics[width=0.9\columnwidth]{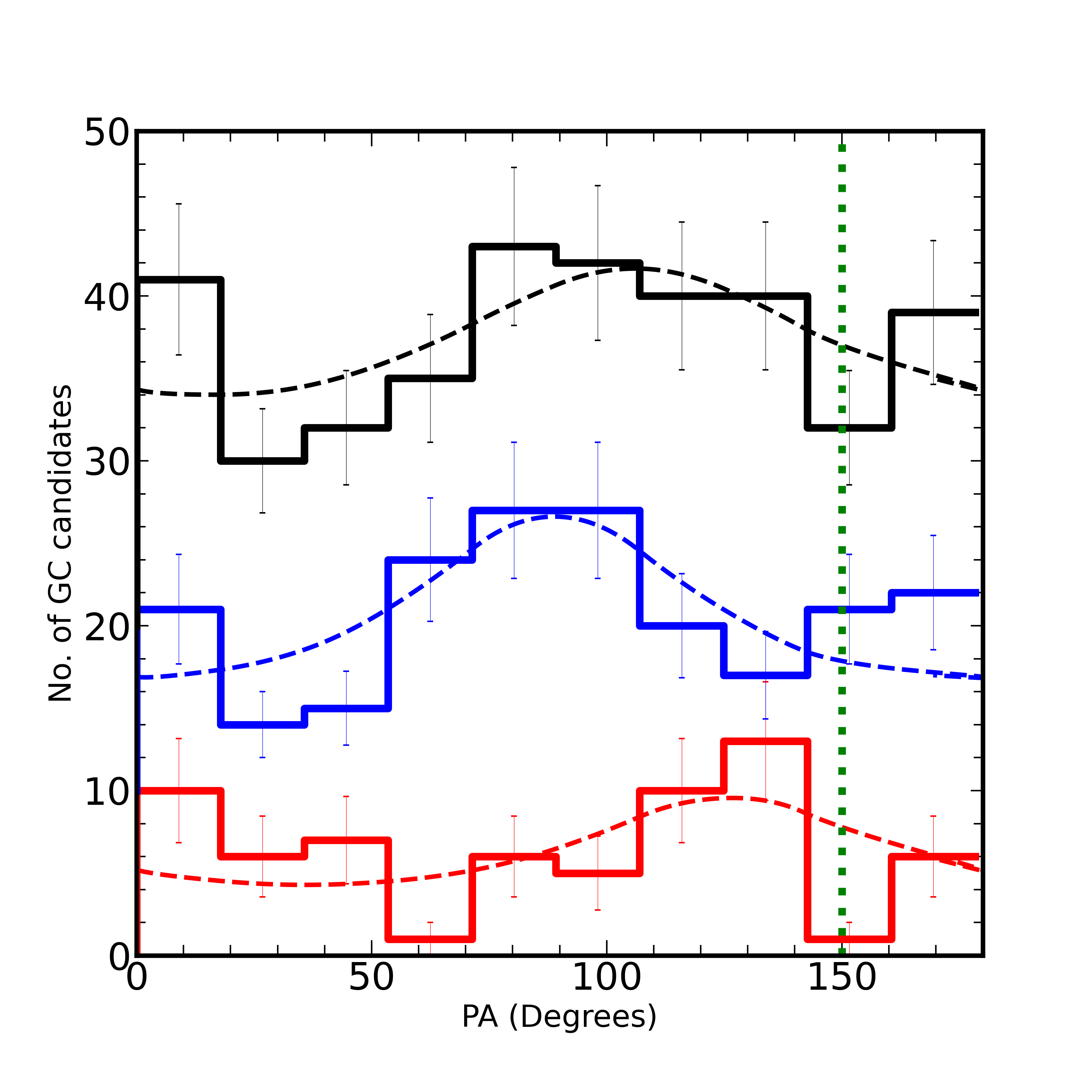}
    \caption{Azimuthal distribution of NGC 4262 GCS. The black, blue, and red histograms represent the azimuthal distribution of the total GCS and the blue and red subpopulations. The dashed lines represent the best fit made on each sample. A vertical dotted green line (at 150$^{\circ}$) indicates the PA of the host galaxy.}
    \label{fig:azim}
\end{figure}

Furthermore, we study the azimuthal distribution of the GCS and their subpopulations. In Figure \ref{fig:azim}, we show the azimuthal distribution of GC candidates.

We estimated the $\epsilon$ and PA of the GCS by following the method outlined in \citet{sreeja2014}
The estimated PA and $\epsilon$ from the fitted profiles are given in Table \ref{tab:PA}, along with the uncertainties, which represent the one-sigma errors. The azimuthal angle of total GCs, and blue and red subpopulations are compared with host galaxy PA. In ETGs, the azimuthal angle of the red subpopulation aligns closely with that of the host galaxy \citep{2013Wang,2018Cantiello,2019ko}. In this case, both blue and red subpopulations show different azimuthal distributions with respect to the host galaxy. However, it is noticeable that the red subpopulation aligns more closely with the host galaxy's PA than the blue subpopulation. In terms of ellipticity, the red subpopulation ($\epsilon$ = 0.6 $\pm$ 0.2) and the host galaxy ($\epsilon$ = 0.13 $\pm$ 0.003) are incomparable, though. In the case of NGC 4546 \citep{2020Escudero} the PA obtained for the red subpopulation differs significantly from the values obtained for the stellar light of the galaxy. They mentioned that this misalignment could be another indication of the violent past of this galaxy.

\begin{table}
\caption{ The position angle and ellipticity values obtained for the total GCs, blue, and red subpopulations are given here. The values for the host galaxy are obtained from the surface photometry analysis (section \ref{sec:surface_PHOTOMETRY}).}
\begin{tabular}{|l|l|l|}
\hline
Type      & PA (Degrees)       & Ellipticity \\ \hline
Galaxy    & 150 $\pm$ 7          & 0.13 $\pm$ 0.003     \\ 
Total GCs & 106 $\pm$ 18     & 0.5 $\pm$ 0.1        \\ 
Blue GCs  & 89 $\pm$ 12      & 0.6 $\pm$ 0.1        \\ 
Red GCs   & 126.5 $\pm$ 20       & 0.6 $\pm$ 0.2        \\ \hline
\end{tabular}
\label{tab:PA}
\end{table}

\begin{figure}
    \centering
    \includegraphics[width=\columnwidth]{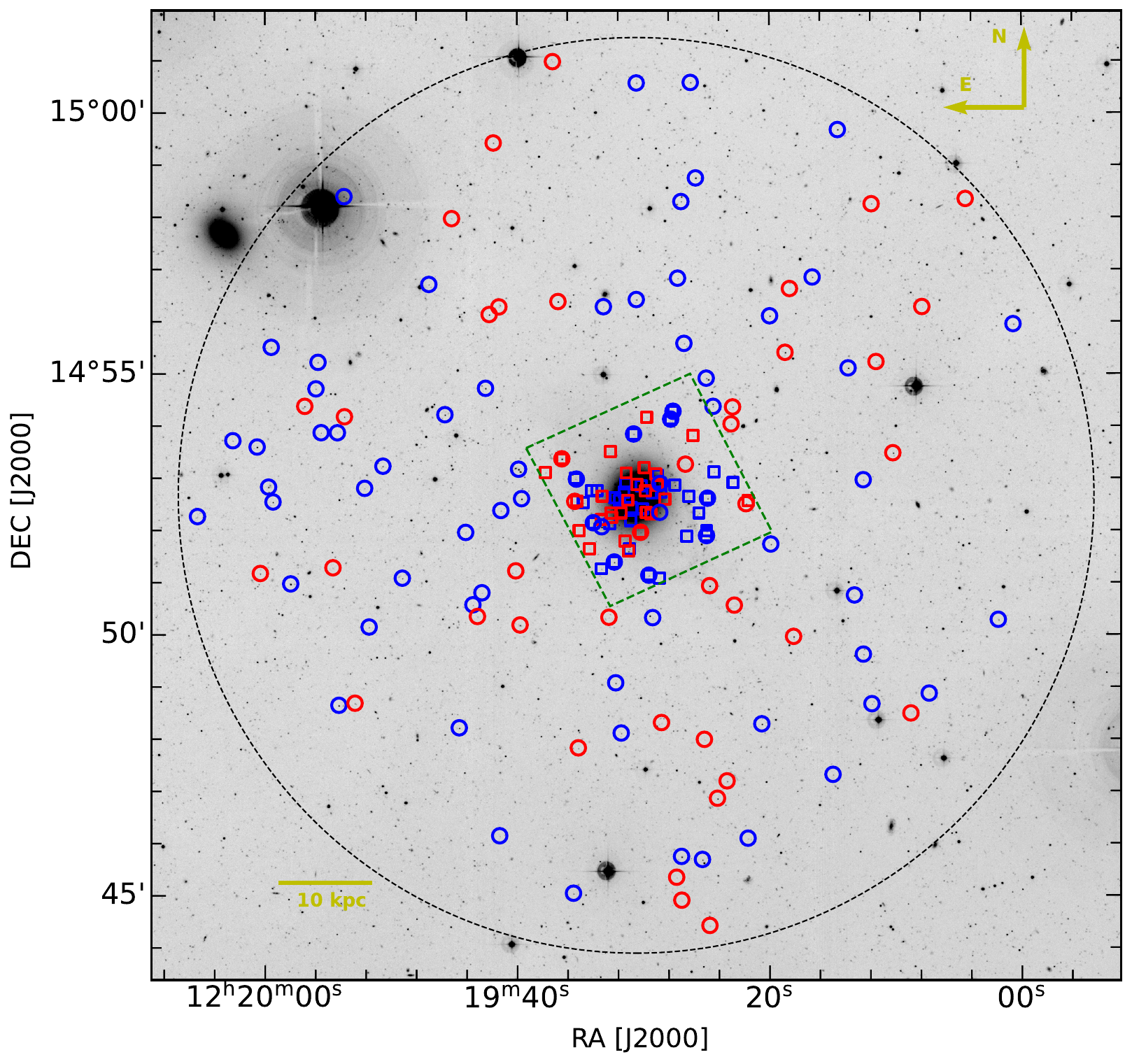}
    \caption{The CFHT optical u-band image of NGC 4262  with a field of view of 18.5\arcmin x 18.5\arcmin. The green dotted line delineates the region observed by HST ACS. The black dotted circle indicates the extent of the GCS. Blue and red colors represent the blue and red GCs identified using HST (open squares) and CFHT (open circles). A horizontal solid yellow line in the bottom left corner indicates the image scale of 10 kpc.}
    \label{fig:GC_u_ovp}
\end{figure}

\begin{figure}
    \centering
    \includegraphics[width=1\columnwidth]{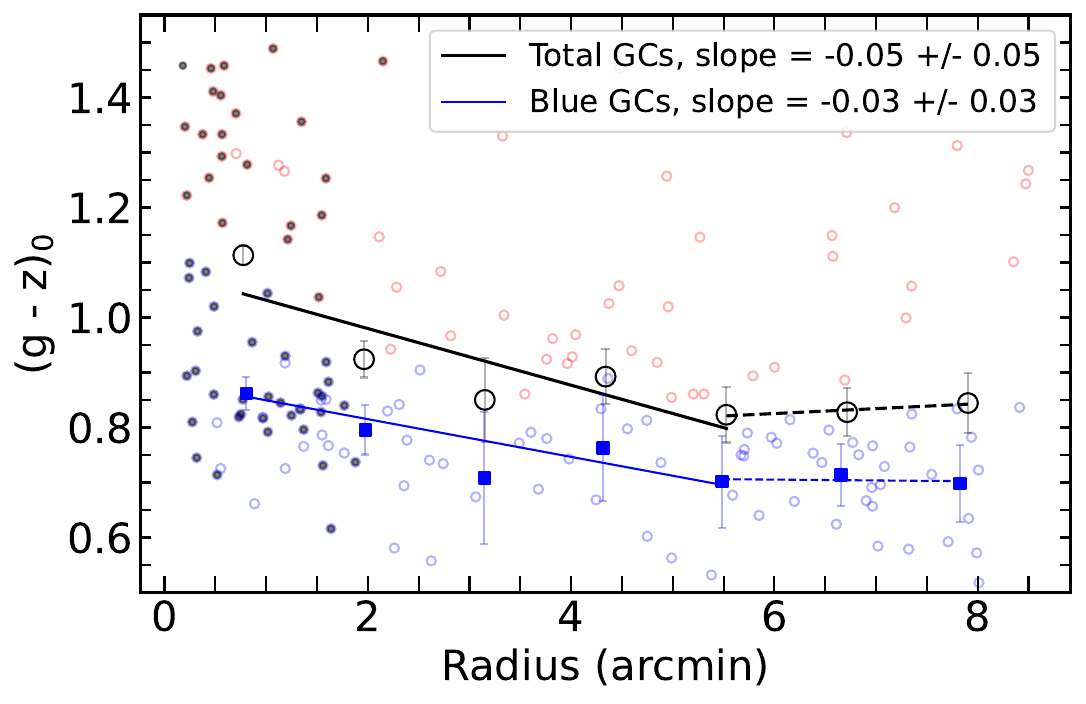}
    \caption{Radial color distribution for the GC system of NGC 4262. The blue and red GCs are shown as small blue and red circles, incorporating data from both HST (filled circles) and CFHT (open circles). The average color separations with errors between the two subpopulations are shown in black open circles. The solid black and blue lines represent the best fit for the total and blue GCs within 5.5\arcmin. The total GCs show a strong color gradient of -0.05 $\pm$ 0.01  within the radius 5.5\arcmin,  transitioning to a flattened color profile beyond this radius. A similar trend has been observed in the blue subpopulations. }
    \label{fig:rad_col}
\end{figure}
The spatial distributions of GCs around NGC 4262 are shown in Figure \ref{fig:GC_u_ovp}, with the blue and red populations represented by blue and red colors, respectively. Additionally, we studied the radial color distribution of GCS of NGC 4262. We employed the methodology outlined in \citet{2011forbe} and \citet{sreeja2014}. Figure \ref{fig:rad_col} presents the radial color distribution for the GC system of NGC 4262, where the blue and red GCs are depicted as small blue and red circles, respectively. The separation between the two subpopulations is determined using a moving mean color in the $g - z$ color within 8$\arcmin$. This running mean value separates red and blue GCs in each radial bin. Then, the mean color for both subpopulations is determined in each radial bin. The blue-filled squares represent the corresponding mean colors and errors for the blue GC population within 8$\arcmin$. However, due to insufficient numbers of red GCs, we could not obtain a best-fitted line within the red population. We observed a color gradient of -0.05 $\pm$ 0.01 in the total population of GCs within a radius of 5.5\arcmin. Beyond this radius, we observed a transition in the color gradient, leading to a flattened color profile. A similar color gradient was noted in the blue subpopulations (-0.03 $\pm$ 0.03), while due to insufficient numbers of red GCs, we were unable to discern any significant color gradient within the red populations. Numerous studies have reported similar transitions in color profiles occurring at various effective radii. Past these points, the profiles tend to flatten out, signifying the absence of continued color gradients \citep{2011forbe,sreeja2014,2021Taylor,2024Caso}. This result also heightens the likelihood of past interactions within NGC 4262, with these interactions being reflected in the properties of its GCS.

\subsubsection{Specific Frequency and total number}
The specific frequency ($S_{N}$) of a galaxy is the total number of GCs in a galaxy per unit host galaxy luminosity \citep{1981Harrisandvanderbergh}.
The given relation defines the value of $S_{N}$,
\begin{equation}
S_{N} = N_{GC} 10^{0.4(M_{V}^{T}+15)}
\end{equation}
The parameter $N_{GC}$ (the total number of GCs) is estimated from the surface density distribution of GCS following the method outlined by \citet{sreeja2014}. $M_{V}^{T}$ is the total absolute V band magnitude of the galaxy. The total absolute magnitude of NGC 4262 obtained from \citet{2013Harris} is -19.5 $\pm$ 0.2 mag. Thus, we determined the total number of GCs in the galaxy is 266 $\pm$ 16, and its specific frequency is 4.2 $\pm$ 0.8.

\begin{figure}
    \centering
    \includegraphics[width=0.9\columnwidth]{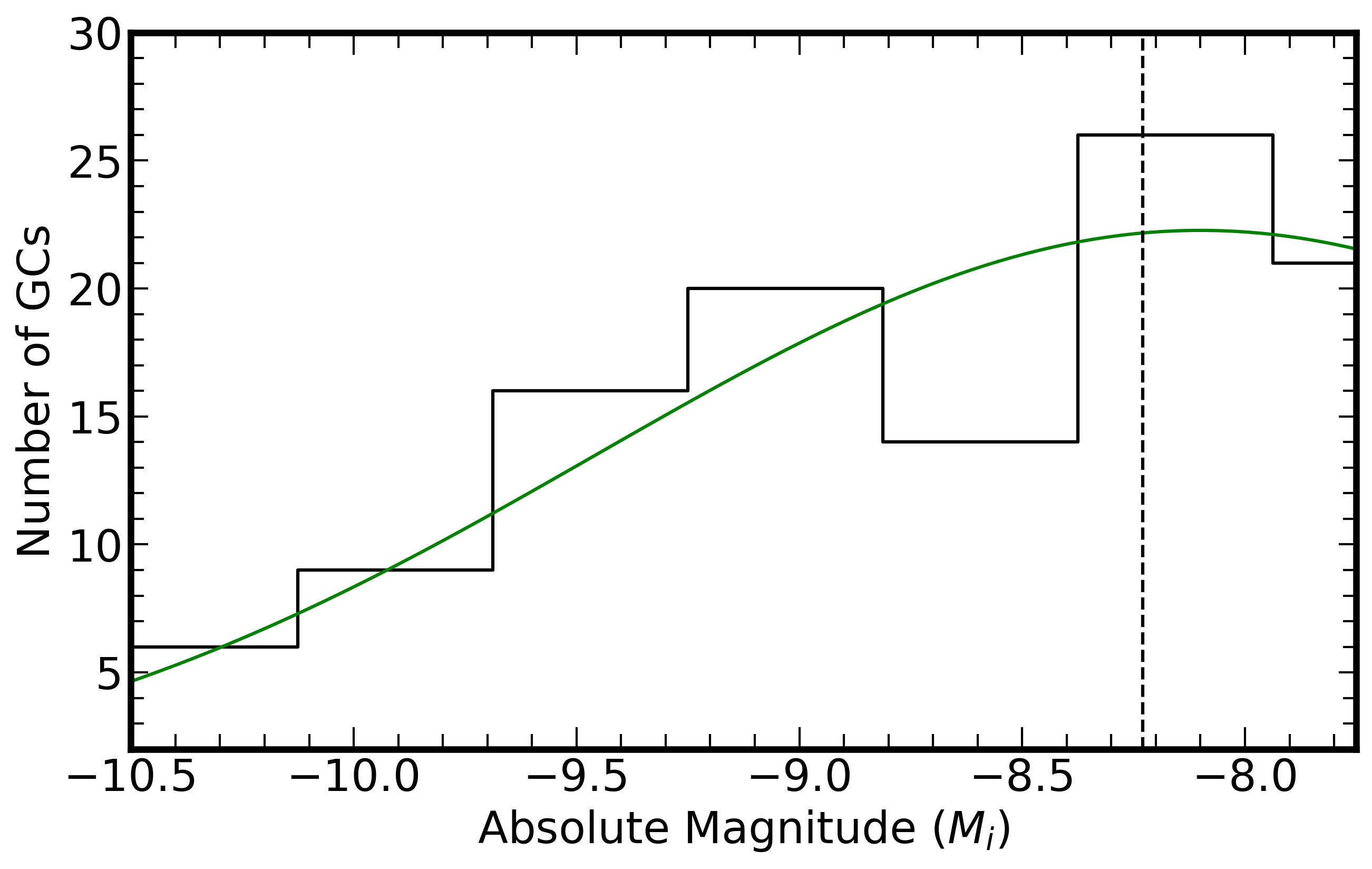}
    \caption{The luminosity function for the GC candidates associated with NGC 4262. The black line represents the i-band absolute magnitude distribution. The green curve corresponds to the Gaussian fit performed. The vertical dotted line corresponds to the turnover magnitude ($M_{i}$ = -8.23 mag). The mean value obtained from the Gaussian fit is -8.1 ± 0.5 mag.}
    \label{fig:GCLF}
\end{figure}

In addition, we have obtained the GC Luminosity Function (GCLF), which quantifies the number of objects per unit magnitude \citep{1993Secker_Harris,2014HarrisGCLF}. Figure \ref{fig:GCLF} represents the GCLF of NGC 4262. We fitted a Gaussian function to the luminosity function, yielding a mean value of -8.1 ± 0.5 mag. Furthermore, the luminosity function enables us to estimate the percentage of the GC population detected in our images. By integrating the Gaussian fit obtained using the parameters up to our magnitude limit (Mi=-7.8 mag), we determine that our detected population comprises 59 ± 19 percent of the total. This further provides the total population as 191 ± 62, which falls within the error limit of the number of GCs estimated from the surface density profile.

In addition, we also evaluated the proportion of a galaxy's baryonic mass attributed to GCs, a metric known as the specific mass \citep[S${_M}$,][]{2008peng,2010Georgiev}. This baryonic mass encompasses the sum of the stellar mass, denoted as $M_{*}$, and the mass of the neutral atomic hydrogen gas, denoted as M$_{HI}$.
\begin{equation}
S{_M} = 100 M_{GCS}/(M_{*}+M_{HI})
\end{equation}
M$_{GCS}$ is the total mass of the GCS. The M$_{HI}$ of NGC 4262 is 4.8 x 10$^{8}$ $M_\odot $, obtained from \citet{2022Yu_HImass}. To calculate the total mass of GCS, we use a mean GC mass of 2.4 x 10$^{5}$ $M_\odot $ \citep{2008Spitler,durell_ngvs2014,2016duncan}. The total mass of the GCS (M$_{GCS}$) in NGC 4262 is estimated to be 6.38 x 10$^{7}$ $M_\odot $ and its $S_M$ is 0.23 $\pm$ 0.01. The estimated values of $N_{GC}$, $S_{N}$ and $S_M$ of NGC 4262 in this study is high value when compared to \citet{2008peng} ($N_{GC}$ =100 $\pm$ 31, $S_{N}$ =  1.70 ± 0.53 and $S_M$ = 0.11 $\pm$ 0.03). It is important to mention that the values for NGC 4262 were derived by \citet{2008peng} through data from the HST ACS survey, which has a smaller field of view than the CFHT. When compared to other lenticular galaxies with similar masses and environments \citep{2015Caso,2019ko}, NGC 4262 exhibits a relatively high $S_{N}$ and $S_{M}$ value. We also noted the similarity of $S_{N}$ value of NGC 4262 with elliptical galaxies \citep[2 < $S_{N}$ < 10;][]{1981Harrisandvanderbergh}.
Higher $S_{N}$ and $S_{M}$ values are typically found in low/high-mass ETGs that have experienced multiple past interactions \citep{2010Georgiev,2013Harris}. This observation suggests that NGC 4262 might exhibit similar characteristics to those of elliptical galaxies, and we can expect a massive halo according to its GCS.

\subsection{Global characteristics of GCS}
\subsubsection{GCS and the dark matter halo of NGC 4262}
To understand more about the GCS properties with respect to the halo of the PRG, we checked the theoretical connections of GCS properties with its halo mass (baryonic and dark mass). Halo masses are commonly defined by the overdensity of the halo with respect to the critical density of the Universe. We defined the halo as a spherical region within the average density $\approx$ 200 times the critical density at the respective redshift.
We estimated the virial mass, M$_{200}$ = 3.08 x 10$^{12}$ $M_\odot $, using the relation mentioned below \cite[eqn: 7.39]{Schneider_peter2015}.
\begin{equation}
M = V_{200}^{3}/10GH(z)
\label{eqn:m200}
\end{equation}
where the Hubble function is given by,
\begin{equation}
H^{2}(z)=H_{0}^{2}[\Omega _{r}(1+z)^{4}+\Omega _{m}(1+z)^{3}+\Omega _{\Lambda }]
\label{eqn:hub}
\end{equation}

\begin{figure}
    \centering
    \includegraphics[width=0.8\columnwidth]{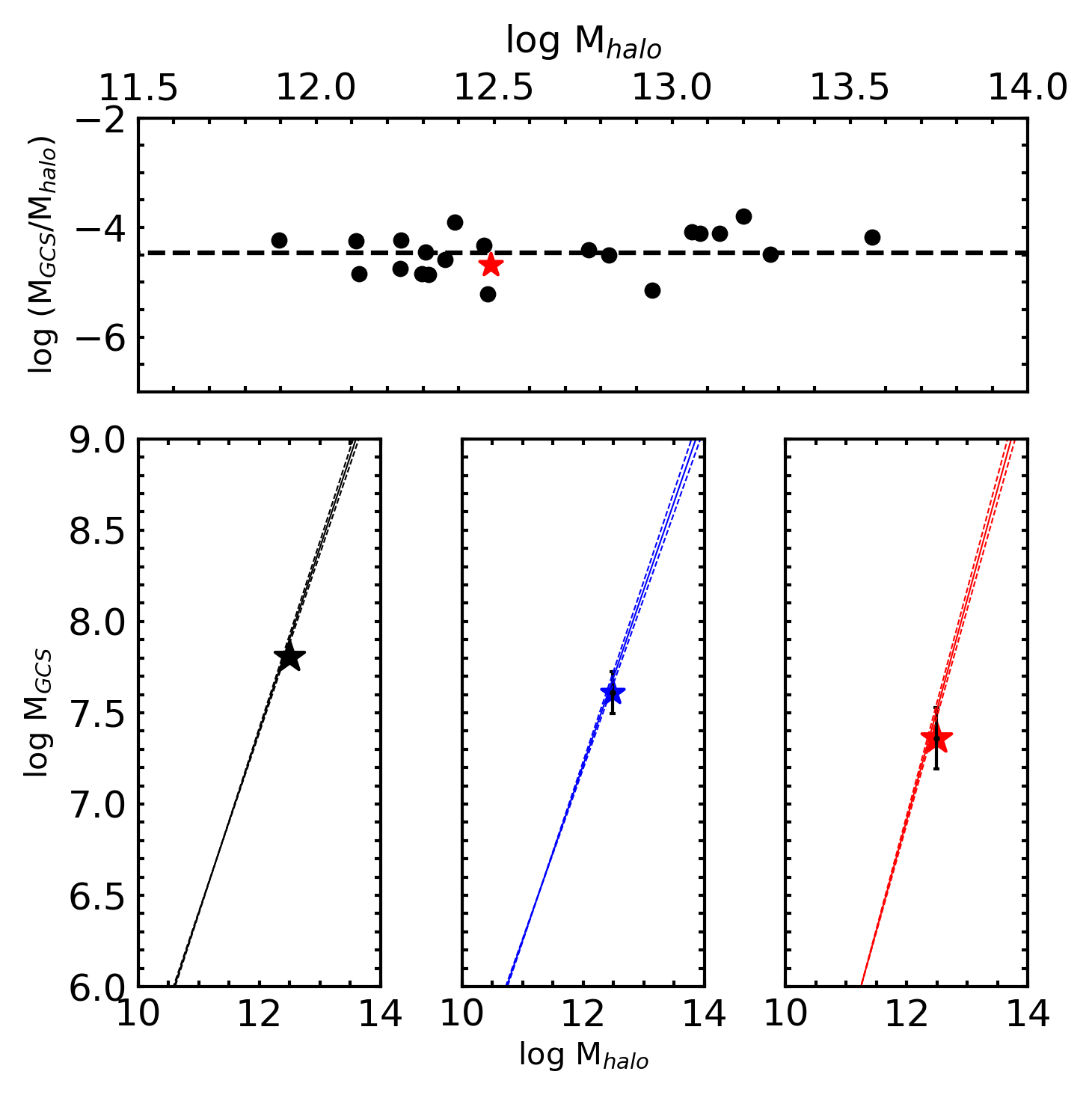}
    \caption{Upper panel: The logarithmic ratio of the mass of GCS and halo mass vs. log halo mass is shown in the Figure. A red star represents NGC 4262, and other black scatter points are galaxies obtained from \citet{2018Hudson}. The dotted line represented the log $\eta$ = −4.54 \citep{2017Harris}. Lower panel: Halo mass relation with GC mass of total GCs, blue, and red subpopulations are represented from left to right, respectively. Solid lines show the linear relations obtained from \citet{2015Harris_hudson}, and the dashed lines are the upper and lower limits for the same. The star marker represents the total GC mass of the galaxy NGC 4262. Our observation indicates that the GCS mass of NGC 4262 remains consistent with the expected halo mass relation, showing no significant deviation.}
    \label{fig:DM_eta}
\end{figure}

Here, we assumed the virial velocity as the maximum circular velocity obtained using HI data for NGC 4262 \cite[210 $\pm$ 10 km/s,][]{Buson2011}. In the previous studies by  \citet{2015Harris_hudson,2017Harris}, a correlation has been established between the mass of the GCS and the mean halo mass (M$_{200}$ + M$_{*}$ + M$_{HI}$). They provided a constant value for the mass ratio, denoted as $\eta$, which is approximately 2.9 × 10$^{-5}$. The mass ratio estimated for the galaxy NGC 4262 in this study is  2.1 × 10 $^{-5}$. The upper panel of Figure. \ref{fig:DM_eta} shows the position of NGC 4262 in the relation between halo mass and GCS mass. Similarly, the lower panels of Figure \ref{fig:DM_eta} represent the  M$_{halo}$ - M$_{GCS}$ relation for total GCs, blue, and red subpopulations. The linear relations for total GCs, blue, and red GC subpopulations are obtained from \citet{2015Harris_hudson}. Our observations indicate that the GCS mass of NGC 4262 does not deviate significantly from the expected halo mass relation. \par

\begin{figure}
\centering
\includegraphics[width=0.9\columnwidth]{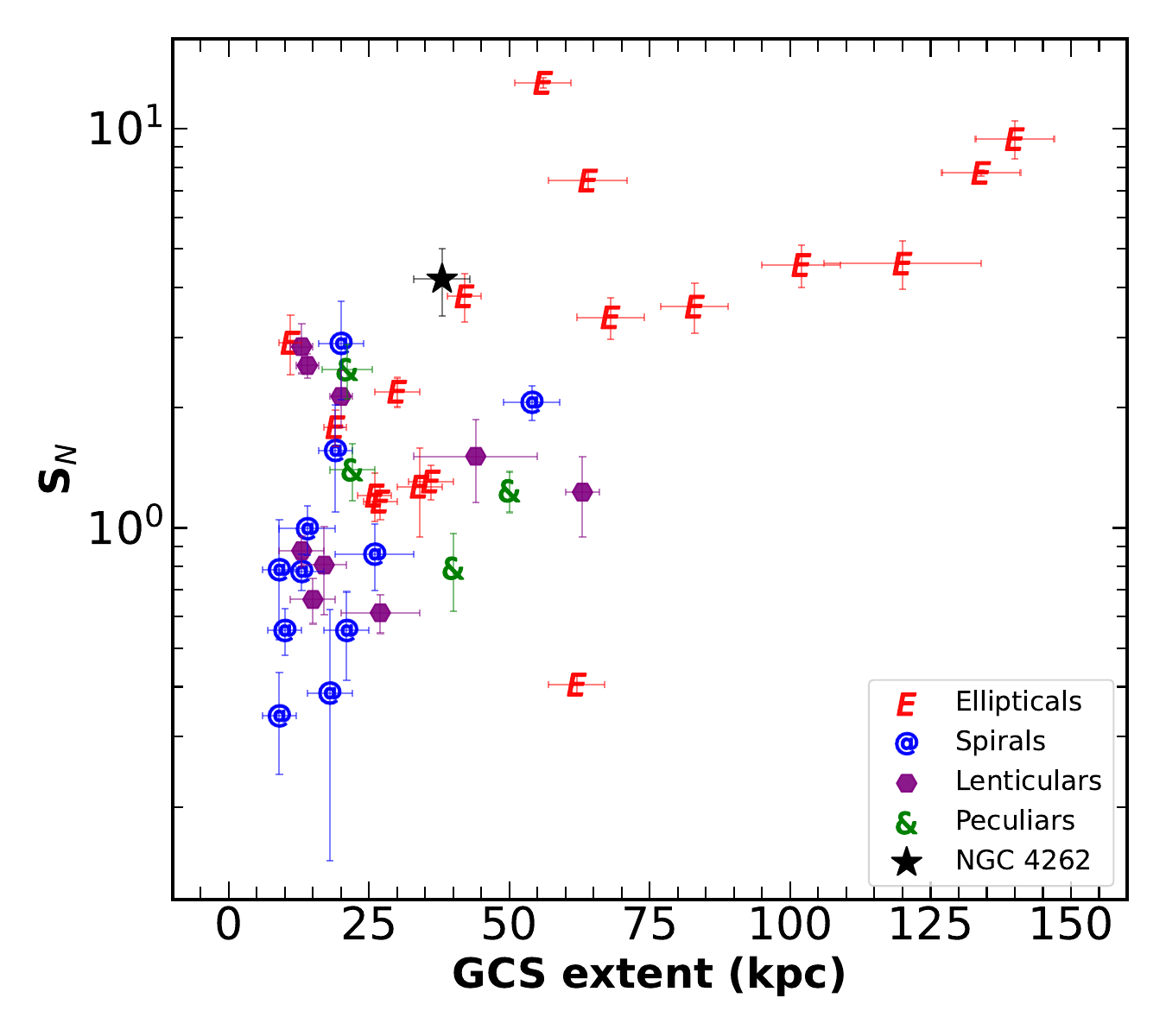}
\caption{GCS extent versus the specific frequency of all 46 galaxies, including NGC 4262, are shown here. NGC 4262 is marked as a black-filled star and compared with other types of galaxies. NGC 4262 inhabits a position within the GCS properties that lies between spiral and elliptical galaxies, with closer proximity to ellipticals. }
\label{fig:global1}
\end{figure}

\begin{table*}
\caption{GCS parameters are listed in the order of galaxy name, distance, absolute V band magnitude, the logarithmic value of stellar mass, the extent of the GCS, the total number of GCs, the ratio of the number of blue GCs and red GCs, effective radius of the host galaxy, and references: a.\citet{2022Escudero}, b.\citet{2015Caso}, c.\citet{2017Lim}, d.\citet{2017Bassinoandcaso} and  e.\citet{2006Sikkema,2015Salinas}}
\scalebox{0.9}{
\begin{tabular}{lllllllll} \hline
NGC & D (Mpc) & Mv (mag) & log(M*) ($M_\odot $) & $GCS_{ext}$ (kpc) & $N_{GC}$ & $N_{BGC}/N_{RGC}$ & $R_{e} $ (kpc) &References\\ \hline
4262        & 15.42           & -19.51           & 10.44                & 38 $\pm$  5                       & 266 $\pm$  16         & 1.8                & 0.9    & \textbf{This work}              \\
4382         & 17.88           & -22.25           & 11.347               & 22 $\pm$  4                       & 1216 $\pm$  82       & -                  & 7.48 & a                          \\
4753         & 23.6            & -22.3            & 10.93                & 50                             & 1030 $\pm$  120       & -                  & 1.69 & b                         \\
474          & 29.51           & -20.97           & 11.078               & 21.1 $\pm$  4.5                   & 609 $\pm$  94         & -                  & 4.82  & c                        \\
3610         & 35              & −22              & 10.6                 & 40                             & 500 $\pm$  110        & 1.5                & 2.62   & d                       \\
2865         & 37.84           & -21.5            & 10.8                 & -                              & 410 $\pm$  8          & 1.63               & 2.63 & e                         \\ \hline
\end{tabular}}
\label{tab:peculiars}
\end{table*}

\begin{figure}
\centering
\includegraphics[width=0.9\columnwidth]{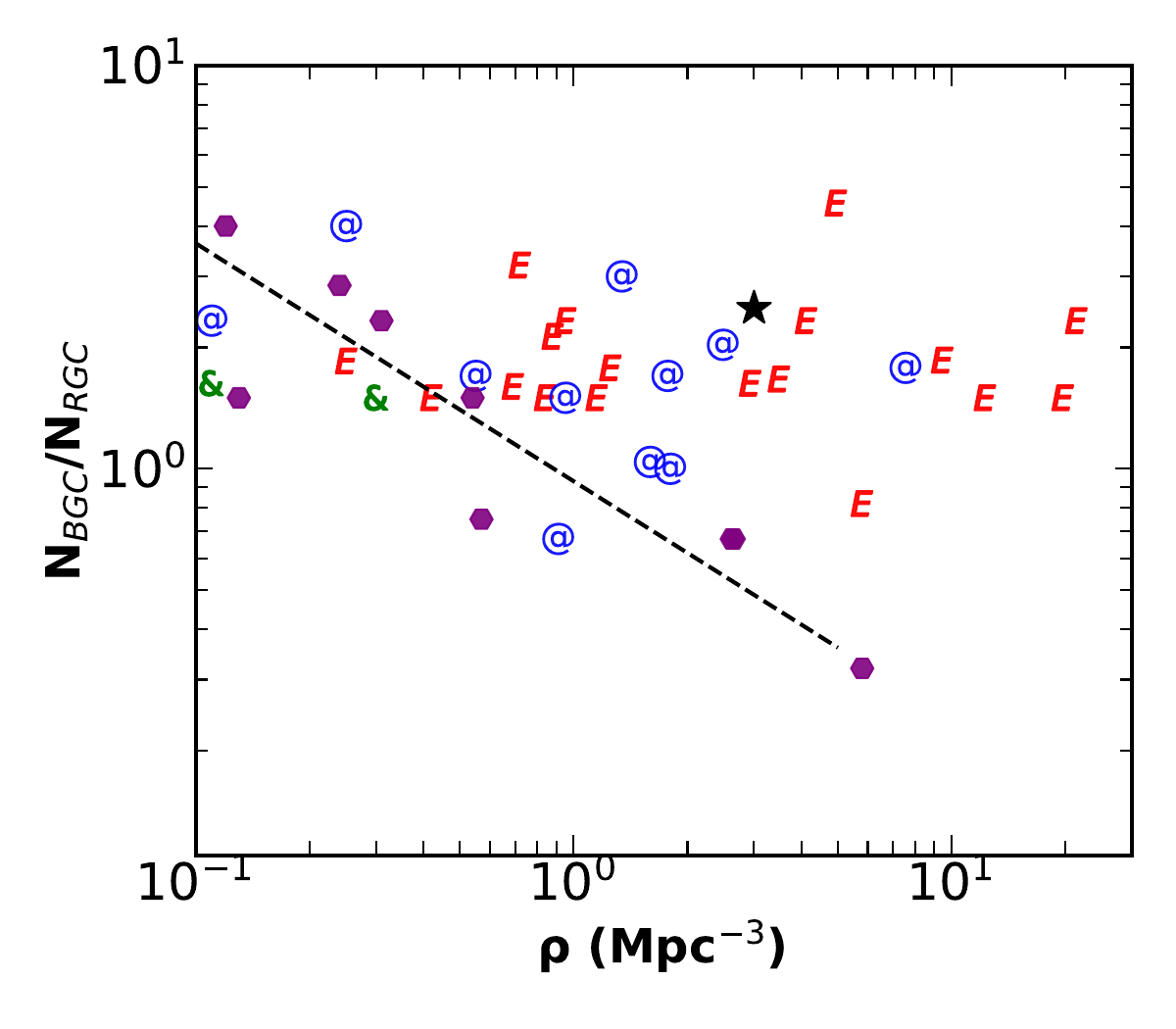}
\caption{This plot gives the number ratio of blue to red GCs versus the density of the environment of different types of galaxies. The symbols in the Figure are the same as in Fig. \ref{fig:global1}. The dotted line represents the relation for lenticular galaxies obtained from \citet{sreeja2014}. This supports the notion that the GCS properties of NGC 4262 are closer to elliptical galaxies than lenticular galaxies.}
\label{fig:global2}
\end{figure}

\subsubsection{Global relations and the GCS of NGC 4262}
\label{section:correlations}
Previous studies \citep[][and reference therein]{brodie2006,2007Rhode,2016Kartha,2019caso,2022debortoli} have considerably explored the properties of GCS in various morphological types of galaxies, except for PRGs. These investigations have uncovered and examined many links between the characteristics of GCS. However, due to the unique nature of PRGs, the understanding of GCS properties in these systems remains relatively unexplored. In this section, we explore the PRG NGC 4262 with other types of galaxies as a function of several GCS properties. The GCS data from \citet{sreeja2014} is utilized as the reference sample. They have provided 40 galaxy samples with GCS properties, and the selection criteria for the sample are explained in \citet[][see section 4]{sreeja2014}. Along with that, we have included five other peculiar galaxies (NGC 4382, NGC 4753,  NGC 474,  NGC 3610, and NGC 2865), and we obtained their GCS properties from previous studies \citep{2006Sikkema,2015Salinas,2015Caso,2017Lim,2017Bassinoandcaso,2022Escudero}. These galaxies share a common trait: they've undergone merger events or show signs of recent accretion. 
GCS of the selected five galaxies is well studied using wide-field imaging and shows different peculiarities in their morphology. The fundamental parameters and GCS properties are provided in Table \ref{tab:peculiars}. \par

In Figure \ref{fig:global1}, the GCS properties of galaxies belonging to different Hubble types are compared with NGC 4262. Based on the GCS properties of various morphological types of galaxies, we observed NGC 4262 possesses a similar orientation to the GCS properties of peculiars. Also, NGC 4262 and peculiar galaxies are located in a region between spiral and elliptical galaxies. Since the host galaxy of NGC 4262 belongs to a lenticular type, the GCS properties should also follow global relations for lenticulars. To address that, we have checked the relation obtained from \citet{sreeja2014} with the ratio of blue to red GCs and the density of the environment. Figure \ref{fig:global2} shows the density of the environment of different types of galaxies obtained from \citet{1988Tully} with the ratio of blue to red GCs. In the Figure, we found NGC 4262 does not follow the trend obtained for the lenticular galaxies.

\section{Discussion}
\label{sec:discussion}
This study extensively investigated the GCS properties of the PRG NGC 4262 using the wide-field optical images obtained from the Canada-France-Hawaii Telescope. Notably, we present the first optical image of NGC 4262 featuring an optically faint ring component (see Figure \ref{fig:igu_color}). We conducted the surface photometric analysis of NGC 4262 and given in Section \ref{sec:surface_PHOTOMETRY}. The optical color profiles revealed the presence of a polar ring structure located beyond a radius of 50". The variation in color signifies the presence of distinct stellar populations within the host and ring components \citep{2024akr}. \par

More importantly, this study represents the first wide-field investigation and analysis of the GCS of a PRG. By thorough photometric analysis, we identified the GCS of PRG NGC 4262. We analyzed the radial density distribution of GCS using Sérsic profiles in section \ref{sec:radialprofile}. From the Sérsic fits, the extent of NGC 4262 GCS is estimated as 38 $\pm$ 5 kpc. The estimated extent agrees with the relation obtained from \citet{sreeja2014}. One of the most studied properties of GCS is its color distribution. GCS of most galaxies was found to be bimodal with blue and red subpopulations \citep{Peng2006,2015Harris_hudson,Hagrisrhode2014,2020debrotoli}. The blue and red subpopulations are believed to correspond to an earlier, metal-poor phase and a later, metal-rich phase of star formation, respectively. From the GMM analysis, we confirmed a bimodality in the GCS of NGC 4262. In the bimodality, our observation indicates that the total GCS consists of 64\% blue subpopulation and 36\% red subpopulation, signifying a dominance of blue GCs by 28\% over red GCs. \citet{2008peng} have studied Virgo cluster ETGs and obtained a red GC fraction relation as a function of host galaxy luminosity. We estimated the fraction of red GCs in NGC 4262 as 0.36, which aligns well with the relation established by \citet{2008peng}.  \par

The results in the azimuthal angle distribution of GCs around the galaxy are intriguing. In ETGs, the host galaxy's position angle and ellipticity match the red subpopulation \citep{2013Wang}. Here, in the case of a PRG, both red and blue subpopulations are not aligned along the galaxy light's PA and are not in good agreement with the ellipticity of the host galaxy. Despite the position angle for the red population placing it closer to the host galaxy than the blue population, the properties of host galaxies and red GCs tend to align with each other in other morphological types of galaxies \citep{2007Rhode,2016Kartha}. Based on the observed results concerning the color gradient of the GCs across the radius of NGC 4262, we can infer the occurrence of a past accretion event, supported by several studies such as \citet{2011forbe}, \citet{2017Caso}, and \citet{2024Caso}. The transition from a color gradient in a galaxy's GCs to a more uniform distribution may indicate a shift from a phase dominated by dissipative processes, such as gaseous collapse or gas-fed flows, to dissipationless mechanisms, such as the accretion of low-mass galaxies \citep{2011forbe,2009Naab,2010Oser,2024Caso}. This alteration in the color gradient of GCs likely mirrors the underlying evolutionary processes that have shaped the galaxy's assembly history. This interpretation finds qualitative alignment with the case of NGC 4262. Based on these findings, it is plausible to suggest that a recent interaction leading to the formation of the ring around NGC 4262, could elucidate the observed deviations in the radial color gradient, as well as the radial and azimuthal distribution of the GCS.

Regarding the total number of GCs, specific frequency, and specific mass of NGC 4262,  the estimated values are higher than those from \citet{2008peng}. Given that the field of view of ACS covers a smaller region than that of NGVS. \citet{2001vandenbergh,2002Beasley,2013Toini,2014Li_hui} explained the formation of blue and red GCs triggered by galaxy interaction. Also, the accretion history of the galaxy is responsible for a significant portion of its GCs \citep{2011forbe}. According to the studies conducted by \citet{2010Georgiev} and \citet{2013Harris}, it has been observed that the relationships of $S_{N}$ and $S_{M}$ exhibit a U-shaped pattern. In this pattern, elevated values of $S_{N}$ and $S_{M}$ tend to appear in both low and high-mass ETGs that have undergone numerous previous interactions. On the other hand, galaxies with intermediate masses tend to exhibit baseline values of $S_{N}$ and $S_{M}$, which are approximately 1 and 0.1, respectively. NGC 4262 is situated in an environment that appears to be less dense. However, it is possible that NGC 4262 had neighboring galaxies during its early stages, leading to multiple interactions that influenced its evolution.

In addition, \citet{2003Iodice} and \citet{2015Moiseev} have discussed the dark matter halo in PRGs and its interesting nature using different simulations. There are various approaches to linking the properties of different Hubble types of galaxies to those of their halos. One of the best methods is the GCS and its properties with the host galaxy \citep{2017Forbes,2018Hudson,2020Bastian}. Our discussion on the connection between the characteristics of GCS and the dark matter halo in PRG NGC 4262 is being presented for the first time. \citet{2014Khoperskov}  investigated the spatial distribution of the dark matter halo surrounding NGC 4262, using advanced modeling techniques. They found that simple halo models could not match the observed data. Instead, they proposed a complex model with a variable shape halo. This model suggests the presence of an oblate halo within the optical radius of the galaxy, gradually flattening towards the perpendicular plane. The axial ratio of the halo varies, being smaller at the inner part of the galaxy and larger at its extent. Thus, the halo of NGC 4262 exhibits an unusual characteristic-- the shape of the dark matter distribution undergoes significant variations with changing radii. From our study, we observed that the interactions had not made any changes in the mass of the GCS. However, the spatial and azimuthal properties within the GCS show remarkable variations, which may be attributed to the peculiar shape of the halo of NGC 4262.\par

The global scaling relations between the host galaxy and GCS in different morphological types of galaxies are well explored \citep[references therein]{sreeja2014,2015Liu,2013Harris,2021alamokarla,2007Rhode}. \citet{sreeja2014,2016Kartha} studied the GCS properties of elliptical, lenticular, and spiral galaxies samples. We included five peculiar galaxies exhibiting signatures of recent interaction in the catalog. Subsequently, a comparison was made between the GCS properties of NGC 4262 and those peculiar galaxies (Figure \ref{fig:global1}). We observed that most of the GCS properties of NGC 4262 possess a similar orientation to the GCS properties of peculiar galaxies. Moreover, we found NGC 4262 and peculiar galaxies located in a transition region between spiral and elliptical galaxies. This interpretation might link the connection between the global evolution scenario of galaxies. Even in the transition region, the high $S_{N}$ of NGC 4262 holds a position close to ellipticals. This suggests that the GCS properties of PRGs are more relative to the ETGs than LTGs. Since the host galaxy of NGC 4262 is lenticular, the GCS properties are expected to follow the lenticular galaxy's global relations. But in Figure \ref{fig:global2}, we observed NGC 4262 showing major variations from the lenticular galaxies trend. Based on the observed GCS properties as mentioned in section \ref{section:correlations}, it appears probable that NGC 4262 could transform into an elliptical galaxy and is currently in a transitional state.

\section{Summary}
\label{sec:summary}
Presenting an optical image of NGC 4262 revealing a faint ring structure, we introduce the first wide-field study of the GCS of a PRG.
\begin{itemize}
    \item We studied the surface brightness distribution of NGC 4262 by employing Sérsic profiles. We observed a bar structure within a 20" radius and a polar ring structure after a 50" radius from the galaxy's center.

    \item A comprehensive investigation has been conducted on the GCS of the PRG NGC 4262. We estimated the extent of NGC 4262 GCS as 38 $\pm$ 5 kpc, with a total of 266 $\pm$ 16 GCs. We also estimated the $S_{N}$ and $S_{M}$ of the galaxy as 4.2 $\pm$ 0.8 and 0.23 $\pm$ 0.01, respectively. The higher value of $S_{N}$ and $S_{M}$ support that the galaxy NGC 4262 has undergone past interactions. 
    
    \item The radial distributions of the subpopulations are fitted with Sérsic profiles. The red subpopulation dominates at the galaxy's center, and the blue subpopulation prevails at a higher radius. From the Sérsic index values we observed the red subpopulation aligns more with the galaxy light distribution than the blue populations. This result is also supported by the position angle of GCS.

    \item Neither the red nor the blue subpopulation aligns with the ellipticity of the host galaxy. The slight deviation observed in the radial and azimuthal angle distribution of the GCS with the host galaxy could be explained by the past interactions of NGC 4262.

    \item Analysis of the radial distribution of GCS in NGC 4262 reveals a color gradient of -0.05 ± 0.01 within 5.5 arcminutes, transitioning to a flattened profile beyond this radius. This suggests past interactions within NGC 4262 and a shift from dissipative to dissipationless mechanisms in its GCS evolutionary history, potentially influenced by recent accretion events.

    \item This study also presents the first results of the GCS and dark matter halo in PRG NGC 4262. An intermediate massive halo was observed for NGC 4262, and the GCS-halo mass ratio was estimated to be 2.1 × 10$^{-5}$. The GCS mass relative to halo mass aligns with the established relations observed in ETGs.

    \item Based on the properties of the GCS, it was noted that both NGC 4262 and the five peculiar galaxies are located in a transition region between spiral and elliptical galaxies. The high specific frequency of NGC 4262 holds a position close to ellipticals, suggesting that the GCS properties of PRGs are more similar to those of elliptical galaxies than spiral galaxies. All these results indicate the probability of NGC 4262 becoming an elliptical galaxy and currently being in a transition state based on its GCS properties. These findings provide insights into the global evolution scenario of galaxies.
    
\end{itemize}
This study provides valuable insights into the structure and properties of PRG NGC 4262 based on its GCS. It is the first comprehensive investigation of the GCS and dark matter halo in NGC 4262, contributing to our understanding of PRGs and their evolutionary processes. Given that this study is considered a pilot study, it opens the possibility of investigating a more extensive sample of PRGs and analyzing their GCS properties to understand the PRGs and their place in galaxy evolution.

\section{Acknowledgement}
AKR and SSK want to acknowledge the financial support from CHRIST (Deemed to be University, Bangalore) through the SEED money project (No: SMSS-2220, 12/2022 ). AKR and SSK express sincere gratitude to Prof Eric Peng for generously sharing the NGVS data and for his invaluable contribution and support throughout the project.
AKR thanks Ujjwal  Krishnan, Robin Thomas, Arun Roy, and Ashish Devaraj for their valuable comments and discussions on the manuscript.  We thank the Center for Research, CHRIST (Deemed to be University), for all their support during this work. AKR, SSK, and BM acknowledge the facility support from the FIST program of DST (SR/FST/PS-I/2022/208). SSK acknowledges the financial support from the Indian Space Research Organisation (ISRO) under the AstroSat archival data utilization program (No. DS-2B-13013(2)/6/2019). This publication uses the data from the  CFHT-NGVS. We gratefully thank all the individuals involved in the various teams for supporting the project from the early stages of the design and observations. This research has used the NASA/IPAC Extragalactic Database (NED), funded by the National Aeronautics and Space Administration and operated by the California Institute of Technology.

\section*{Data Availability}
The data underlying this article will be shared on reasonable request to the corresponding author.

\bibliographystyle{mnras}
\bibliography{example}




\bsp	
\label{lastpage}
\end{document}